\documentclass[aps,pra,reprint]{revtex4-2}%\documentclass[10pt,preprint,twoside]{osajnl}

\usepackage{amsmath,amssymb,graphicx,physics}
\usepackage{sanitize-umlaut}
\usepackage{array}
\usepackage[caption=false]{subfig}
\usepackage{bm}
\usepackage{amsfonts}
\usepackage{epsfig,epstopdf}
\usepackage{subfig}
\usepackage[hyphenbreaks]{breakurl}
\usepackage{mathtools}

\def\cN{{\cal N}}
\def\Tr{{\rm Tr}}

\def\la{\left\langle}
\def\ra{\right\rangle}

\def\br{{\bf r}}

\def\bL{{\bf L}}

\def\bC{{\bf C}}
\def\bL{{\bf L}}

\def\bu{{\bf u}}

\def\bI{{\bf I}}

\def\bG{{\bf G}}

\def\bH{{\bf H}}

\def\be{\begin{equation}}
\def\ee{\end{equation}}
\def\ba{\begin{align}}
\def\nn{\nonumber\\}

\def\defeq{\buildrel \rm def \over =}
\def\bR{{\bf R}}
\def\ll{\left\langle}

\def\b0{{\bf 0}}

\def\cL{{\cal L}}
\def\la{\left\langle}
\def\ra{\right\rangle}

\def\br{{\bf r}}

\def\bL{{\bf L}}
\def\bu{{\bf u}}

\def\bC{{\bf C}}
\def\bL{{\bf L}}

\def\bI{{\bf I}}

\def\bG{{\bf G}}

\def\bH{{\bf H}}

\def\cO{{\cal O}}
\def\cL{{\cal L}}
\def\cI{{\cal I}}

\def\Tr{{\rm Tr\ }}

\def\be{\begin{equation}}
\def\ee{\end{equation}}
\def\ba{\begin{align}}
\def\nn{\nonumber\\}
\def\la{\langle}
\def\ra{\rangle}

\def\defeq{\buildrel \rm def \over =}
\def\bR{{\bf R}}

\def\cS{{\cal S}}

\def\br{{\bf r}}
\def\brp{{\bf r'}}

\def\pmu{\partial_\mu}
\def\pnu{\partial_\nu}

\def\ha{{\rm h.a.}}
\def\ll{<\!\!<}
\begin{document}

\title{Quantum Limits on Localizing Point Objects against a Uniformly Bright Disk}

\author{Sudhakar Prasad}
\email{prasa132@umn.edu}
%\author{Zhixian Yu}
%\author[1,*]{Sudhakar Prasad}
\affiliation{School of Physics
and Astronomy, University of Minnesota, Minneapolis, MN 55455}
\altaffiliation{Also in Department of Physics and Astronomy, University of New Mexico, Albuquerque, NM 87131}

%\affil[*]{sprasad@unm.edu}
\date{\today}

%\doi{\url{http://dx.doi.org/10.1364/ol.XX.XXXXXX}}

\begin{abstract}

We calculate the quantum Fisher information (QFI) for estimating, using a circular imaging aperture, the two-dimensional location of a point source against a uniformly bright disk of known center and radius in the ideal photon-counting limit. We present both a  perturbative calculation of the QFI in powers of the background-to-source brightness ratio and a numerically exact calculation of the QFI in the eigen-basis of the one-photon density operator. A related problem of the quantum limit on estimating the location of a small-area brightness hole in an otherwise uniformly bright disk, a problem of potential interest to the extrasolar planet detection community, is also treated perturbatively in powers of the ratio of the areas of the hole and the background disk. We then numerically evaluate the Cram\'er-Rao lower bound (CRB) for wavefront projections in three separate bases, those comprised of Zernike, Fourier-Bessel and localized point-source modes, for unbiased estimation of the two position coordinates of the point source and of the brightness hole center, respectively, for the two problems. By comparing these CRBs with the corresponding quantum-limited minimum error variances, given by inverting the QFI matrix, and with the CRBs associated with direct imaging, we assess the maximum efficiency of these wavefront projections in performing such estimations.   
\end{abstract}

%\setboolean{displaycopyright}{true}
%activate for two-column option
\vspace{-1cm}

%\begin{center}
%\today
%\end{center}

\maketitle
%\thispagestyle{fancy}
%\ifthenelse{\boolean{shortarticle}}{\abscontent}{}

\section{Introduction}\label{sec:intro}

A number of high-resolution imaging applications involve optical localization of point objects against a luminous background. Single-molecule localization microscopy (SMLM) \cite{Khater20,Lelek21} typically concerns the localization of individual labels using their flourescence signal photons against finite background illumination presented by other molecules located outside the field of view of an imaging frame. A less typical but important application of SMLM involves a superresolution shadow imaging (SUSHI) technique \cite{Tonnesen18} for visualizing the extra-celluar space (ECS) in the brain in which the tissue cells moving through the ECS cast shadows against the fluorescence from a dye-labeled ECS background.  Industrial machinery and aircraft components may begin to fail due to the development of tiny cracks and point imperfections \cite{Pernick80,Fan21} from repetitive mechanical motion and general material fatigue. Early interventions based on illuminating parts suspected of imminent failure to detect such imperfections seem to be greatly desirable. In the astronomical domain, there is overwhelming interest in exoplanet (EP) detection \cite{Fischer14}, using instruments like TESS \cite{TESS15} for detection of EPs transiting across their parent star disks \cite{Charbonneau00} and the Gemini Planet Imager (GPI) \cite{Macintosh14} for direct optical detection of EPs. 

Different operating conditions apply to the different applications. The background illumination could be relatively faint, as for the typical SMLM problem, or bright as for SUSHI and the EP detection problem, and the source localization may involve sub-diffractive scales in the extreme faintness limit, as for the latter problem. Cameras that count and correct wavefronts at the level of individual photons, such as the EMCCD based wavefront sensor upgrade of GPI 2.0 \cite{GPI2}, must be utilized to collect remote optical data from such source-background pairs in order to have the best chance of overcoming these challenging conditions. 

We present here the ultimate estimation theoretic limit, that furnished by quantum Fisher information (QFI) \cite{Helstrom76,Holevo82,Toth14, Liu20, D-D20}, on the variance of estimation of the two-dimensional (2D) location of a point source against a fixed, uniformly bright, circular background disk in the ideal photon-counting limit. We also address a closely related problem of a brightness hole in an otherwise uniformly bright disk source, which might describe the problem of an EP partially occulting the disk of its parent star. 

A previous calculation of the quantum-limited minimum error variances in estimating the separation of an asymmetrically bright point-source pair, which might describe a non-overlapping EP-star pair, appeared in Ref.~\cite{Prasad20a}.  An asymptotic error analysis of a related problem of {\it detection}, rather than {\em estimation} of the separation, of a non-overlapping EP-star system based on quantum binary hypothesis testing and an interferometric realization of such quantum-limited detection in the laboratory have recently been presented \cite{Huang21,Zanforlin22}. Unlike the present work, however, all these previous studies assumed that both the star and planet were themselves unresolved, or point-like. 

The two problems of localizing an unresolved source and a tiny hole inside a finite-radius, uniformly bright background disk require slightly different methods of mathematical analysis, but they both draw from our previous eigenfunction-eigenvalue based numerically exact approach to compute QFI for estimating the spatial parameters of extended sources such as circular and elliptical disks \cite{Prasad20b}. Both these problems are also well positioned for an accurate perturbative treatment in certain limits. We develop a new perturbative analysis based on directly evaluating the symmetric logarithmic derivatives (SLDs), in terms of which the QFI is defined, in the limits of a bright source against a faint background and a tiny hole in the brightness of a larger disk. The perturbative parameters in the two cases are evidently the ratio of the background to source irradiances and the ratio of the hole area to the disk area, respectively. We will validate the results of the perturbative analysis against a numerically exact treatment for the first problem. A perturbative analysis alone will suffice for the second problem. Some preliminary results for the first problem appeared in Ref.~\cite{Prasad22}.

We will also consider the use of projections \cite{Tsang16,YuPrasad18,PrasadYu19} of the photon wavefront that is emitted by the point source - background disk combination and the hole containing luminous disk into an orthonormal basis of modes as a way of estimating the 2D location of the source and the hole, respectively. We will assess the theoretical efficiency of three different modal projection bases, namely the Zernike basis \cite{Noll76}, Fourier-Bessel basis \cite{Watson95,Lebedev72}, and a novel localized point-source basis, and of direct imaging to perform this localization task by computing the corresponding Fisher information (FI) matrix \cite{Kay93}. We will then compare the lowest possible estimation error variance, the Cram\'er-Rao bound (CRB), given by the diagonal elements of the inverse of the FI matrix, for each localization protocol with the ultimate quantum-mechanically admissible lower bound, the quantum CRB (QCRB), obtained by inverting the QFI matrix. 

\section{Localization of a Point Source against a Uniform Background Disk}

\begin{figure}
\centerline{
\includegraphics[width=0.3\textwidth]{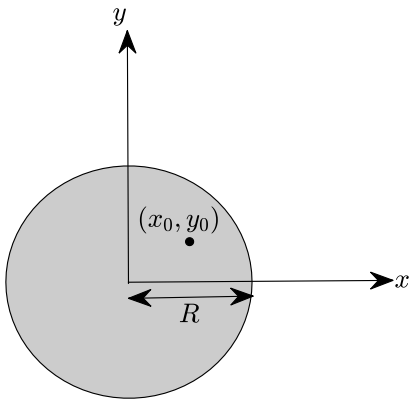}}
\caption{A point source at position $(x_0,y_0)$ within a uniformly bright disk background of radius $R$}
\end{figure}
Figure 1 depicts our first problem, that of estimating the coordinates, $(x_0,y_0)$, of the position of a point source within an otherwise uniformly bright disk of fixed radius, $R$, and center, which we choose to be the origin of the coordinate system. Let the ratio of the integrated background and point-source brightnesses be $b:1-b$. Assuming that the entire disk is in the field of view of a well corrected imaging system with a circular exit pupil, we may express the density operator of a single photon emitted from either the point source or the background disk and then captured by the imager as \cite{Prasad20b}
\ba
\label{SPDO1}
\varrho=&(1-b)\ketbra{K_0}{K_0}+{b\over\pi R^2}\int_B dA\,\ketbra{K_\br}{K_\br}\nn
         =&(1-b)\varrho_0+b\varrho_B,
\end{align}
in which $\ket{K_0}$ denotes the normalized state vector of the photon emitted by the point source located at position vector $\br_0=(x_0,y_0)$, $\ket{K_\br}$ the state of the photon emitted by a point $\br$ in the background disk, and 
\be
\label{SPDOsb}
\varrho_0=\ketbra{K_0}{K_0}\ {\rm and}\  \varrho_B={1\over \pi R^2}\int_B dA\,\ketbra{K_\br}{K_\br}
\ee
are the states of the photon emitted by the point source and the background disk, respectively. The subscript $B$ on the integral sign denotes integration over the background disk area. The mixed-state single-photon density operator (SPDO) of Eq.~(\ref{SPDO1}) for emission by either the point source or a background-disk point reflects the perfectly incoherent emission of the photon from these sources. 

The normalized wavefunction of the photon emanated from point $\br$ of the source and subsequently received by the circular exit pupil of the imager may be expressed as
\be
\label{wavefunction}
\braket{\bu}{K_\br} ={1\over\sqrt{\pi}}\Theta_P(\bu)\,\exp(-i2\pi\bu\cdot\br),
\ee
where $\Theta_P(\bu)$ is the indicator function of the clear pupil, taking the values 1 inside and 0 outside, and whose radius, $R_0$, has been scaled to be 1 by means of the variable transformation, $\bu\to R_0\bu$. The position vector $\br$ of a source point is also expressed in scaled units, with the scale factor being the characteristic diffraction parameter, $w_d=\bar\lambda z_I/R_0$, for mean imaging wavelength $\bar\lambda$ and image-plane distance from the exit pupil $z_I$. The latter scaling amounts to the variable transformation, $\br\to w_d\br$, in the image plane. The background disk radius, $R$, is also expressed in the same units by dividing the physical disk radius in the image plane by $w_d$. Expressing transverse separations in the image plane in units of $w_d$ has the immediate benefit that all so-scaled separations smaller than 1 are in the sub-diffractive, superresolution-imaging domain while those larger than 1 are in the superdiffractive, conventionally resolvable domain. 

Typical values of $w_d$ are of the order of a few $\mu$m both for a high-numerical-aperture, high-magnification microscope \cite{Lelek21} and an EP transit-detection telescope \cite{Charbonneau00} operating in visible light. The SMLM problem when idealized as the localization of a point source against a luminous background disk entails a background brightness of about 50-100 photons per pixel over 50-70 pixels of a single-molecule image containing mean signal photon numbers of order 10$^4$ \cite{Yildiz03,Cabriel19}. That corresponds to a typical value for $b$ in the range of 0.2-0.4.
%for which a perturbative calculation that we present next may not be adequate.

\subsection{A Perturbative Computation of QFI for $b \ll 1$}

A perturbative approach to computing the QFI is warranted in two opposite limits for this problem, namely when $b\ll1$ and $1-b\ll1$, corresponding to the limits of weak and strong background illumination. We will defer addressing the latter limit until after we have introduced the second problem, that of localizing an unresolved brightness hole in the background disk, to which it is trivially isomorphic. The perturbative approach we follow here is novel but simpler and better adapted than the recently discussed Frech\'et-derivative-based operator-expansion method \cite{Grace22} for evaluating the QFI.
 
The QFI is the supremum of Fisher information (FI) over all possible measurements that can be made on a quantum system, corresponding to all possible positive-operator valued measures (POVMs) on its Hilbert space. For a state described by the density operator $\varrho$ that depends differentiably on the parameters, $\theta_1,\ldots,\theta_n$, the QFI is a symmetric real matrix with elements,
\ba
\label{QFIdef}
H_{\mu\nu} = &{1\over 2}\left[\Tr\left(\varrho\cL_\mu\cL_\nu\right)+\Tr\left(\varrho\cL_\nu\cL_\mu\right)\right]\nn
                  = &\Tr\left(\pmu\varrho\,\cL_\nu\right)=\Tr\left(\pnu\varrho\,\cL_\mu\right),
\end{align}
in which $\pmu$ is a shorthand notation for the partial derivative $\pdv{}{\theta_\mu}$, further understood to operate on the single quantity immediately following it. The operator $\cL_\nu$ denotes the SLD of the density operator with respect to the parameter $\theta_\nu$, and is defined by the relation
\be
\label{SLDdef}
\pnu\varrho={1\over 2}\left(\cL_\nu\varrho+\varrho\cL_\nu\right).
\ee
The second line of Eq.~(\ref{QFIdef}) results from using the cyclic property of the operator trace and substituting the defining relation (\ref{SLDdef}) for either $\cL_\mu$ or $\cL_\nu$ into the first line of Eq.~(\ref{QFIdef}).

For the SPDO (\ref{SPDO1}), only its $\varrho_0$ part contains the parameters to be estimated, namely the coordinates $x_0,y_0$ of the point-source position. We may therefore express the corresponding QFI using the second line of Eq.~(\ref{QFIdef}) and the relation (\ref{SLDdef}) defining the SLD as 
\ba
\label{QFISLD}
H_{\mu\nu}=&(1-b)\, \Tr\left(\pmu\varrho_0\, \cL_\nu\right)\nn
                =&(1-b)\left(\pmu\bra{K_0}\cL_\nu\ket{K_0}+{\rm c.c.}\right);\nn
\pnu\varrho_0=&{1\over 2}\left[\cL_\nu(\varrho_0+\alpha\varrho_B)+{\rm h.a.}\right],
\end{align}
where the symbols c.c.~and h.a.~denote the complex conjugate and Hermitian adjoint, respectively, of the terms inside the parentheses or brackets, and $\alpha$ defined as
\be
\label{alpha}
\alpha={b\over 1-b}
\ee
will be treated as a small parameter in the faint-background limit. To arrive at the second identity in the first of Eqs.~(\ref{QFISLD}), we used the fact that $\pmu\varrho_0=\pmu\ketbra{K_0}{K_0}+\ket{K_0}\pmu\bra{K_0}$ and the property of the trace, $\Tr(\ketbra{\Psi}{\Phi}A)=\bra{\Phi}A\ket{\Psi}$. 

We next expand the SLD in powers of $\alpha$ as
\be
\label{SLDpowerseries}
\cL_\nu=\cL_\nu^{(0)}+\alpha \cL_\nu^{(1)}+\alpha^2\cL^{(2)}_\nu+\ldots.
\ee
If we now substitute expansion (\ref{SLDpowerseries}) into the first of Eqs.~(\ref{QFISLD}), we arrive at a perturbative expansion of the QFI:
\be
\label{QFIp}
H_{\mu\nu}=(1-b)\sum_{n=0}^\infty \alpha^nK^{(n)}_{\mu\nu},
\ee
where 
\be
\label{Kn}
K_{\mu\nu}^{(n)}=2\Re \left(\pmu\bra{K_0}\, \cL_\nu^{(n)}\ket{K_0}\right),
\ee
in which Re (Im) denotes the real (imaginary) part of the quantity that follows it. 

A similar substitution into the second of Eqs.~(\ref{QFISLD}), followed by comparing terms of the same power in $\alpha$ on both sides of the resulting equation, yields the following relations for the different orders of the SLD operator:
\ba
\label{SLDp}
\pnu\varrho_0=&{1\over 2}\left(\cL_\nu^{(0)}\varrho_0+{\rm h.a.}\right)\nn
-\left(\cL_\nu^{(n-1)}\varrho_B+{\rm h.a.}\right)=&\left(\cL_\nu^{(n)}\varrho_0+{\rm h.a.}\right),\ n=1,2,\ldots
\end{align}

A perturbative term-by-term evaluation of the QFI (\ref{QFIp}) starts with solving for the zeroth-order SLD, $\cL_\nu^{(0)}$, given by the first of Eqs.~(\ref{SLDp}), which is a Lyapunov matrix equation, in exponential form \cite{Bellman97}, 
\be
\label{Lyapunov}
\cL^{(0)}_\nu=2\lim_{\eta\to 0^+}\int_0^\infty dx \exp(-x\varrho_\eta)\pnu\varrho_0\exp(-x\varrho_\eta),
\ee
where $\varrho_\eta$ is a full-rank extension \cite{Safranek17} of the rank-1 density operator $\varrho_0$, which is defined as
\be
\label{FullRankDO}
\varrho_\eta=(1-\eta)\varrho_0+{\eta\over D}\cI,
\ee
in which $D$ is the Hilbert space dimensionality and $\cI$ is the identity operator in that space. For infinite dimensional spaces, like the one the imaging photon belongs to in the present problem, a spatial discretization of the background disk to which a finite value of $D$ applies may be needed to justify our analysis. The actual value of $D$ turns out to be irrelevant, however, for our subsequent analysis.

For the rank-1 SPDO $\varrho_0=\ketbra{K_0}{K_0}$, we may evaluate integral (\ref{Lyapunov}) and thus $\cL^{(0)}_\nu$ fully. The second of the relations (\ref{SLDp}) has a similar exponential solution, and may also be fully evaluated for an arbitrary finite-order SLD in terms of the next lower-order SLD. Details of these evaluations and a perturbative analysis of the QFI matrix elements to quadratic order in $\alpha$ are presented in Appendix A. To quadratic order, the QFI matrix elements may be expressed as
\be
\label{QFI2orders2}
H_{\mu\nu}=(1-b)\left(K_{\mu\nu}^{(0)}+\alpha K_{\mu\nu}^{(1)}+\alpha^2 K_{\mu\nu}^{(2)}\right),\ \ \mu,\nu=x,y,
\ee
where 
%in which complete expressions for $K^{(0)}_{\mu\nu}$, $K^{(1)}_{\mu\nu}$, and $K^{(2)}_{\mu\nu}$ are given in Eqs.~(\ref{K0}), (\ref{K1}), and (\ref{K2}), respectively, and reproduced below for ease of reference:
\ba
\label{K012}
K_{\mu\nu}^{(0)}=&4\Re\pmu\!\bra{K_0}\pnu\!\ket{K_0};\nn
K^{(1)}_{\mu\nu} =&-4\Re\big[\pmu\!\bra{K_0}\pnu\!\ket{K_0}\mel{K_0}{\varrho_B}{K_0}\nn
                                    &\qquad\qquad+\pmu\!\bra{K_0}\varrho_B\pnu\!\ket{K_0}\big];\nn
K^{(2)}_{\mu\nu}=&4\mel{K_0}{\varrho_B}{K_0}^2\Re (\pmu\!\bra{K_0}\pnu\!\ket{K_0})\nn
                           &+8\mel{K_0}{\varrho_B}{K_0}\Re (\pmu\!\bra{K_0}\varrho_B\pnu\!\ket{K_0})\nn
                           &+4\Re (\pmu\!\bra{K_0}\varrho_B^2\pnu\!\ket{K_0})\nn
                           &+4\Im (\pmu\!\mel{K_0}{\varrho_B}{K_0})\Im(\bra{K_0}\varrho_B\pnu\!\ket{K_0})\nn
                           &+4\Re (\pmu\!\mel{K_0}{\varrho_B}{K_0}\pnu\!\mel{K_0}{\varrho_B}{K_0}).
\end{align}

If one of the coordinate axes is chosen to align with the vector direction of the point source location relative to the center of the background disk, then for a clear circular-aperture imager that we assume here, the invariance of the problem under reflection about that axis implies that every term on the right-hand side (RHS) of expressions (\ref{K012}) must vanish unless $\mu=\nu$. The diagonalization of the QFI matrix, as we will see, is true to all orders in such a coordinate system. Note that for $\mu=\nu$ the first order correction, as given in Eq.~(\ref{K012}), is always negative and thus reduces the zeroth order QFI as the background brightness increases, which is physically sensible. 

%The various matrix elements on the right-hand sides of expressions (\ref{K012}) may be readily evaluated, as shown in Appendix B, using the explicit form (\ref{SPDOsb}) of $\varrho_b$ and the photon wavefunction (\ref{wavefunction}) in aperture coordinates. The final expressions of these matrix elements can be reduced to either a single or double area integral over the background disk, which were then numerically evaluated.   
Another important observation about the perturbative results (\ref{QFI2orders2}) and (\ref{K012}) is that they do not require the knowledge of the eigenvalues and eigenstates of the SPDO, and thus can be readily calculated for arbitrary geometries and brightness disributions of the background against which the point source is to be localized. This is particularly advantageous since, as we will see presently from Figs.~2-5, the perturbative analysis seems to be quite accurate even out to a 1:1 background-to-source irradiance ratio in certain cases. %, at least for the estimation of the azimuthal source coordinate against a disk shaped uniformly illuminated background. 
We defer further discussion of the perturbative results until after we have discussed a numerically exact calculation of the QFI matrix.

\subsection{A Numerically Exact Evaluation}
In a previous paper \cite{Prasad20b}, we calculated the QFI for estimating the radius of a circular disk shaped source by first computing the full sets of non-zero eigenvalues, $\{\lambda_1,\lambda_2,\ldots\}$, and the corresponding eigenstates, $\{\ket{\lambda_1},\ket{\lambda_2},\ldots\}$, of the SPDO and then using the following exact expression for the QFI matrix elements:
\begin{align}
\label{Hmunu1}
H_{\mu\nu}&=\sum_i{4\over \lambda_i}\Re\bra{\lambda_i}\partial_\mu \varrho\partial_\nu \varrho\ket{\lambda_i}\nn
&+2\sum_{i,j}\!\left[{1\over \lambda_i+\lambda_j}\!-\!{1\over \lambda_i}\!-\!{1\over\lambda_j}\right]\nn
&\qquad\times\Re\big(\bra{ \lambda_i}\pmu \varrho\ket{\lambda_j}\bra{\lambda_j}\pnu \varrho\ket{\lambda_i}\big),
\end{align}
in which the $i$ and $j$ sums are over only those eigenstates that have nonzero eigenvalues. The present problem is only slightly more general in that the scene also contains a point source located somewhere in the field of the background disk and one must estimate the source location, rather than the disk radius. The mathematical approach developed in that paper to compute the eigenvalues and eigenstates can therefore be applied here with only slight modifications. Note that other recent, numerically exact, finite-marix based approaches for computing the QFI that do not require first the diagonalization of the SPDO into its orthonormal eigenstates \cite{Safranek18,Fiderer21} are of little value here because the Hilbert space for our problem is intrinsically infinite dimensional.  

We first express SPDO (\ref{SPDO1}) in a fully integral form  as
\be
\label{SPDO1int}
\varrho=\int_B dA\,G(\br)\,\ketbra{K_\br}{K_\br}, 
\ee
where $G(\br)$ is defined using the 2D Dirac $\delta$ function as 
\be
\label{Gdef}
G(\br)\defeq (1-b)\delta^{(2)}(\br-\br_0)+{b\over\pi R^2}\Theta_B(\br)
\ee
and $\Theta_B(\br)$ is the indicator function for the background disk. Each eigenstate of the SPDO may be similarly expressed as a linear combination of states in the support of $\varrho$,
\be
\label{SPDO1eigenstate}
\ket{\lambda}=\int_B dA\,G(\br)\,C_\lambda(\br)\ket{K_\br}.
\ee
Substituting this expansion into the eigenvalue equation, $\varrho\ket{\lambda}=\lambda\ket{\lambda}$, and then comparing coefficients of each state vector $\ket{K_\br}$ on both sides of the resulting equation
yields the following integral equation that the coefficient function $C_\lambda(\br)$ must obey over the disk:
\be
\label{CoeffFnEqn}
\int_B dA' G(\brp) {2J_1(2\pi|\br-\brp|)\over 2\pi |\br-\brp|} C_\lambda(\brp) =\lambda C_\lambda(\br).
\ee
To arrive at Eq.~(\ref{CoeffFnEqn}), we used the fact that the inner product between single-photon states emanating from two different source points $\br$ and $\brp$ and captured by the circular imaging aperture (of unit scaled radius), is simply
\be
\label{KrKrp}
\bra{K_\br}\ket{K_\brp}={2J_1(2\pi|\br-\brp|)\over 2\pi|\br-\brp|}.
\ee 

Let us now consider the Gegenbauer expansion formula \cite{Prasad20b,Watson95},
\ba
\label{BesselAddThm}
{2J_1(2\pi|\br-\brp|)\over 2\pi|\br-\brp|}=2&\sum_{m=0}^\infty\sum_{n=-m,-m+2,\ldots}^m (m+1){J_{m+1}(2\pi r)\over 2\pi r}\nn
&\times{J_{m+1}(2\pi r')\over 2\pi r'}\exp[in(\phi-\phi')],
\end{align}
where $\rho,\phi$ and $\rho',\phi'$ are the polar coordinates of $\br$ and $\brp$, respectively.
Its substitution into the integral equation (\ref{CoeffFnEqn}) immediately shows that the coefficient function, $C_\lambda(\br)$, may be expressed as the generalized Fourier-Bessel sum
\be
\label{CoeffFnBesselSum}
C_\lambda(\br)=\sum_{m,n}(m+1)^{1/2} C_{mn}{J_{m+1}(2\pi r)\over 2\pi r}\exp[in(\phi-\phi_0)],
\ee
in which the sum over $m$ runs over all nonnegative integers and that over $n$ runs between $-m$ and $m$ in steps of 2, just as in Eq.~(\ref{BesselAddThm}). All such double sums henceforth will be understood to have the same ranges. 

By substituting expression (\ref{CoeffFnBesselSum}) for the eigenfunctions along with expression (\ref{Gdef}) for $G$ into Eq.~(\ref{CoeffFnEqn})  and comparing the two sides of the resulting equation term by term, we may easily show that the coefficients $C_{mn}$ obey the eigenvalue equation,
\ba
\label{CmnEqn}
\sum_{m',n'} M_{mn;m'n'} C_{m'n'}=&\lambda C_{mn},\nn
m=0,1,\ldots; &\ n=-m,-m+2,\ldots,m,
\end{align}
in which the four-dimensional array $M$, which can be read off from the resulting area integral over $\brp$, takes the form
\ba
\label{Mmn}
&M_{mn;m'n'} = \sqrt{(m+1)(m'+1)}\nn
&\quad\times \Bigg[{2b\over \pi^2 R^2}\delta_{nn'}\int_0^{R}dr' {J_{m+1}(2\pi r')J_{m'+1}(2\pi r')\over r'}\nn
 &\quad+{(1-b)\over \pi^2 r_0^2}J_{m+1}(2\pi r_0)J_{m'+1}(2\pi r_0)]\Bigg].
 \end{align}

Equation (\ref{Mmn}) becomes a matrix eigenvalue equation if one regards the two pairs of indices, $(m,n)$ and $(m',n')$, as being mapped lexicographically to two single indices. We adopt exactly such a vectorization approach to solve this equation numerically by first truncating the $m,m'$ sums at a finite upper cut-off value that is large compared to $c=2\pi R$ beyond which the Bessel functions inside the radial integral in expression (\ref{Mmn}) become super-exponentially small throughout the radial range $(0,R)$. Further, the integral involving the Bessel functions in Eq.~(\ref{Mmn}) can be evaluated analytically in closed form in terms of other Bessel functions, as we showed in Ref.~\cite{Prasad20b}. We then use the {\em eig} routine in Matlab to calculate the eigenvalues and eigenvectors of the array $M$ and thus the coefficient functions $C_\lambda(\br)$ defined via relation (\ref{CoeffFnBesselSum}).  The eigenvectors generated by {\em eig} have to be properly renormalized to ensure the desired normalization, $\bra{\lambda}\ket{\lambda^\prime}=\delta_{\lambda\lambda^\prime}$. The convergence of the sum of the eigenvalues to 1, which must hold for any density operator, provided a check on the validity of the choice of the upper cut-off of the $m,m'$ sums and on our overall calculations. 

\subsubsection{Reality and Reflection Symmetry of Eigenfunctions}

Since the integral equation (\ref{CoeffFnEqn}) has a real kernel and its eigenvalues are all real, we may choose the eigenfunctions $C_\lambda(\br)$ to be real as well. Since expression (\ref{Mmn}) for the array elements is invariant under a simultaneous sign change of both $n$ and $n'$, the arrays $C_{mn}$ and $C_{m,-n}$, according to Eq.~(\ref{CmnEqn}), are degenerate in their eigenvalues. This implies that the arrays $C_{mn}$ are, or may be chosen to be, either symmetric or antisymmetric in their $n$ index,   
\be
\label{Reality}
C_{mn}=\pm C_{m,-n},\ n=-m,-m+2,\ldots,m,
\ee
and thus the eigenfunctions given by the sum (\ref{CoeffFnBesselSum}) chosen to be real and either even or odd under reflection in the vector joining the point source location to the disk center, {\em i.e.,} under the transformation $(\phi-\phi_0)\to -(\phi-\phi_0)$. We henceforth call these eigenfunctions simply {\em even} and {\em odd} eigenfunctions.

Once we have computed a sufficiently large number of eigenvalues and eigenstates accurately, we substitute them into expression (\ref{Hmunu1}) for the QFI matrix elements and evaluate the sums numerically. In Appendix B, we provide further details of the simplifications, including some useful identities and sum rules already established in Ref.~\cite{Prasad20b}, that are needed to perform the numerical evaluations efficiently. We show that the QFI matrix elements may be expressed as
\be
\label{Hmunu2}
H_{\mu\nu}=4(1-b)^2 \Re \tilde H_{\mu\nu},
\ee
where ${\tilde H}_{\mu\nu}$ denotes the expression,
\ba
\label{Hmunu3}
{\tilde H}_{\mu\nu}& ={1\over 1-b}\pmu\!\bra{K_0}\cI_N\pnu\ket{K_0}\nn
                  +\sum_{i,j\in\cS}&{\lambda_i\lambda_jC_i(\br_0)C_j(\br_0)\bra{\lambda_i}{\pmu}\ket{K_0}\bra{\lambda_j}{\pnu}\ket{K_0}\over\lambda_i+\lambda_j}\nn
+\sum_{i,j\in\cS}&{\lambda_i^2C_i^2(\br_0)\pmu\!\bra{K_0}\ket{\lambda_j}\bra{\lambda_j}{\pnu}\ket{K_0}\over\lambda_i+\lambda_j},
\end{align}     
in which $\cN$ and $\cS$ denote the sets of index values that label the eigenstates in the null and support subspaces, respectively, of the SPDO. The symbol $\cI_N$ denotes the identity operator in the null subspace,
\be
\label{IdentNull}
\cI_N=\cI-\sum_{i\in\cS}\ket{\lambda_i}\bra{\lambda_i}.
\ee

For a full-rank SPDO, the first term on the RHS of Eq.~(\ref{Hmunu3}) vanishes identically. Our detailed evaluations of the first term suggest this to be true to within numerical accuarcy. Thus, only the two double sums contribute to the QFI.  In fact, since $\lambda_iC(\br_0)=\braket{\lambda}{K_0},$ we see that both these double sums are over bilinear products of the overlap integrals between the SPDO eigenstates and the point-source emission state and between the SPDO eigenstates and appropriate derivatives of the point-source emission state. Such bilinear products decrease rapidly with decreasing eigenvalues. Thus, despite the sum of eigenvalues being in the denominator of each term of these double sums, the contributions of these terms decrease rapidly with decreasing eigenvalues.  

All expressions in Eq.~(\ref{Hmunu3}) involve only one class of nontrivial matrix elements in need of numerical evaluation, namely those of form $\bra{\lambda_i}\pmu\ket{K_0}$, which, in view of expression (\ref{SPDO1eigenstate}) for the eigenstates, may be written as the integral,
\ba
\label{lambdai_pmu_K0}
\bra{\lambda_i}\pmu&\ket{K_0}
=\int C_i(\br)\, G(\br)\bra{K_\br}\pmu\ket{K_0} dA\nn
=&{b\over \pi R^2}\int_B C_i(\br)\bra{K_\br}\pmu\ket{K_0} dA\nn
=&{b\over \pi R^2}\int_B C_i(\br)\pmu\braket{K_\br}{K_0} dA\nn     
=&{2b\over \pi R^2}\int_B C_i(\br){J_2(2\pi|\br-\br_0|)\over |\br-\br_0|^2}(x_\mu-x_{0\mu}) dA.
\end{align}
In Eq.~(\ref{lambdai_pmu_K0}) the fact that the $\delta$ function part of $G(\br)$, as defined in Eq.~(\ref{Gdef}), cannot contribute to the integral, since $\bra{K_0}\pmu\ket{K_0}=0$ for either value of $\mu$, was used to arrive at the second equality, while the fact that $\pmu$ operates only on the state $\ket{K_0}$ was used to derive the next equality. A subsequent substitution of the overlap function (\ref{KrKrp}) and use of a simple Bessel-function derivative identity,
\be
\label{Bessel}
{d\over dx}[x^{-n}J_n(x)]=-x^{-n}J_{n+1}(x),
\ee
along with the derivative formula,
\be
\label{derivative}
\pmu |\br-\br_0|={x_\mu-x_{0\mu}\over |\br-\br_0|},
\ee
where $x_\mu,x_{0\mu}$ are the $\mu$th component of $\br,\br_0$, led to the final equality. The area integral (\ref{lambdai_pmu_K0}) over the disk may be calculated quite efficiently by using Matlab's built-in {\em integral2} code. When combined with a highly accurate numerical evaluation of the eigenvalues and eigenvectors, this allowed us to evaluate the QFI matrix elements (\ref{Hmunu3}) quite accurately. 

\subsubsection{Diagonalization and Parameter Compatibility of the QFI Matrix}
Note that the QFI matrix is, in general, off-diagonal since the matrix element (\ref{lambdai_pmu_K0}) does not vanish for either component of the source location vector $\br_0$. It can, however, be diagonalized by a proper choice of coordinate axes, specifically if the $x$ axis is chosen to be along $\br_0$ for which $\phi_0=0$. For this choice, the matrix element $\bra{\lambda_i}\pmu\ket{K_0}$ is nonzero only if either the eigenfunction $C_i(\br)$ is even and $\mu=1$ or $C_i(\br)$ is odd and $\mu=2$. That is because for $\phi_0=0$ the integrand of the disk-area integral in expression (\ref{lambdai_pmu_K0}) is odd under reflection, $\phi\to-\phi$, for the other two possibilities, namely either even eigenfunctions and $\mu=2$ or odd eigenfunctions and $\mu=1$, and thus the $\phi$-integral over its full, symmetric range, $(-\pi,\pi)$, vanishes. Further, since odd eigenfunctions vanish for $\phi=0$, it follows that $C_i(\br_0)=0$ for an odd eigenfunction. Taken together, these two results imply that the first double sum in Eq.~(\ref{Hmunu3}) can only receive contributions from those $i,j$ index values that label the even eigenfunctions and thus $\mu=\nu=1$, while the second double sum there, although not restricted to even eigenfunctions in its $j$ index, still requires that $\mu=\nu$ in order not to vanish identically. In other words, the two double sums add up to 0 unless $\mu=\nu$. The first term on the RHS of Eq.~(\ref{Hmunu3}) was already noted to be vanishingly small to numerical precision. The full QFI matrix is thus diagonal within this choice of coordinate axes. 
%Its third term also has the same character, since $C_i(\br_0)$ and $C_j(\br_0)$ are non-vanishing only for the even eigenfunctions and thus only even eigenfunctions contribute to this double sum. The condition $\mu=\nu$ must again hold for the matrix element product $\bra{\lambda_i}\pmu\ket{K_0}\bra{\lambda_j}\pnu\ket{K_0}$ to be non-zero when only even eigenfunctions are involved.  then only the even $C_i(\br)$ is nonvanishing for $\br=\br_0$. This means that only even eigenfunctions contribute to both

Further, since $\tilde H_{\mu\nu}$ given by Eq.~(\ref{Hmunu3}), whose real part yields the QFI matrix element $H_{\mu\nu}$ according to Eq.~(\ref{Hmunu2}), is already real and vanishing whenever $\mu\neq \nu$ in the coordinate axes aligned with the radial position vector of the point source and its orthogonal direction,  the radial and azimuthal location coordinates are compatible parameters \cite{Ragy16}. In other words, these two source position coordinates can be independently estimated and the error variances of their estimates can asymptotically saturate the QCRBs given by the reciprocals of the diagonal elements of the QFI matrix.  As we will see, this is also true for the second problem discussed later.

Since the diagonalizing axes are oriented along and orthogonal to the source position vector, the reciprocals of the diagonal elements of the diagonal QFI matrix, $\bH^{(0)}$,  are the minimum variances of {\em unbiased} estimation of the radial distance, $r_0$, of the point source from the disk center and of its angular coordinate times $r_0$.  If needed, the QFI matrix, $\bH$, with respect to the original pair of coordinate axes, oriented at angle $-\phi_0$ relative to the diagonalizing axes, is easily obtained by means of a similarity transformation of ${\bf H}^{(0)}$, 
\be
\label{Hxy}
{\bf H}=\bR_0 \bH^{(0)} \bR_0^{\rm T},
\ee
by the rotation matrix, $\bR_0$, that has the form, 
\be
\label{rotmat}
\bR_0 =
\begin{pmatrix}
\cos\phi_0& \sin\phi_0\\ -\sin\phi_0& \cos\phi_0
\end{pmatrix}.
\ee

\subsubsection{Numerical Results and Discussion}
We evaluated $\bH^{(0)}$ by numerically computing the largest 40-50 eigenvalues, associated eigenfunctions, and integrals of kind (\ref{lambdai_pmu_K0}) for the background disk radius taking values out to $R=2$. We checked for the convergence of the QFI sum (\ref{Hmunu3}) by enlarging the truncated set of eigenvalues to include progressively lower eigenvalues until we reached to a high precision of roughly 1 part in 10$^{5}$. At this stage, the sum of the eigenvalues checked out to be equal to 1 to within 1 part in 10$^9$, which was confirmation that we had numerically exhausted the support space of the SPDO over which the sums in expression (\ref{Hmunu3}) are evaluated and thus of the high precision of our overall computation. 

We next evaluated the perturbative expression (\ref{QFI2orders2}) of QFI out to the quadratic order in the background strength parameter, $\alpha$, defined by Eq.~(\ref{alpha}). This required computing the matrix elements in Eq.~(\ref{K012}), which we carried out in the same coordinate system in which the exact QFI matrix is diagonal. In view of the background-disk SPDO, $\varrho_B$, given by Eq.~(\ref{SPDOsb}), these matrix elements take the form,
\ba
\label{PertMatEl}
\mel{K_0}{\varrho_B}{K_0}=&{1\over \pi^3 R^2}\int_B {J_1^2(2\pi |\br-\br_0|)\over |\br-\br_0|^2} dA;\nn
\mel{K_0}{\varrho_B\pmu}{K_0}=&{2\over \pi^2 R^2}\int_B {J_1(2\pi |\br-\br_0|)J_2(2\pi|\br-\br_0|)\over |\br-\br_0|^3}\nn
&\ \qquad\qquad\times (x_\mu-x_{0\mu})\, dA;\nn
\pnu\!\mel{K_0}{\varrho_B\pmu}{K_0}=&\delta_{\mu\nu}{4\over \pi R^2}\int_B {J_2^2(2\pi |\br-\br_0|)\over |\br-\br_0|^4}\nn
&\ \qquad\qquad\times(x_\mu-x_{0\mu})^2 dA;\nn
\pnu\!\mel{K_0}{\varrho_B^2\pmu}{K_0}=&\delta_{\mu\nu}{4\over \pi^3 R^4}\int_B dA\int_B dA' {J_2(2\pi |\br-\br_0|)\over |\br-\br_0|^2}\nn
&\ \times{J_2(2\pi |\brp-\br_0|)\over |\brp-\br_0|^2}{J_1(2\pi|\br-\brp|)\over |\br-\brp|}\nn
&\  \times (x_\mu-x_{0\mu})(x'_\mu-x_{0\mu}),
\end{align}
in which $x_{0\mu}$ is equal to $r_0$ for $\mu=1$ and 0 for $\mu=2$. To derive these expressions, we made use of rather similar mathematical manipulations to those that led to expression (\ref{lambdai_pmu_K0}) for the matrix element $\mel{\lambda_i}{\pmu}{K_0}$. Further, the fact that the last two matrix elements in Eq.~(\ref{PertMatEl}) are nonvanishing only if they are diagonal follows from the invariance of both the background disk and the point source location vector under reflection in the coordinate axis aligned with that vector. 

The single area integrals that determine the first three matrix elements in Eq.~(\ref{PertMatEl}) were accurately computed using the {\em integral2} routine as before. The double area integral in the final matrix element required the use of a multi-dimensional integral code \cite{Hosea22}, which provides excellent computational accuracy and efficiency for such four-dimensional integrals.  

In Fig.~2, we display the variation of the QFI, $H_{11}$, for estimating the radial distance of the point source from the center of the background disk with increasing value of the integrated background brightness level, $b$.   In the absence of the background ($b=0$), this radial-distance QFI has its maximum value of $4\pi^2$, which we had noted earlier in Ref.~\cite{YuPrasad18}. With increasing value of $b$, the point source becomes progressively fainter in the ratio $(1-b):b$ relative to the background, resulting in an almost linear decrease of the radial-distance QFI, particularly for values of $R$ larger than 1. The numerically exact results, shown by solid curves, are hardly affected by the actual radial distance, $r_0$, of the point source from the disk center, as we see for two different values of $r_0$, namely $0.2R$ and $R$ corresponding to the source being close to the center and at the edge of the disk. This is particularly true for values of $R$ larger than 1. The second-order perturbative results, which we show by dashed curves, are always quite accurate for $b\ll1$, but for $R$ smaller than 1, those results are quite accurate even out to $b=0.5$ for which the perturbation parameter, $\alpha$, is quite large at 1. The departure at even moderately large values of $b$ from the exact results is quite pronounced, however, when $R$ is larger than 1, as we see from the figure for $R=2$.
\begin{figure}
\centerline{
\includegraphics[width=0.55\textwidth]{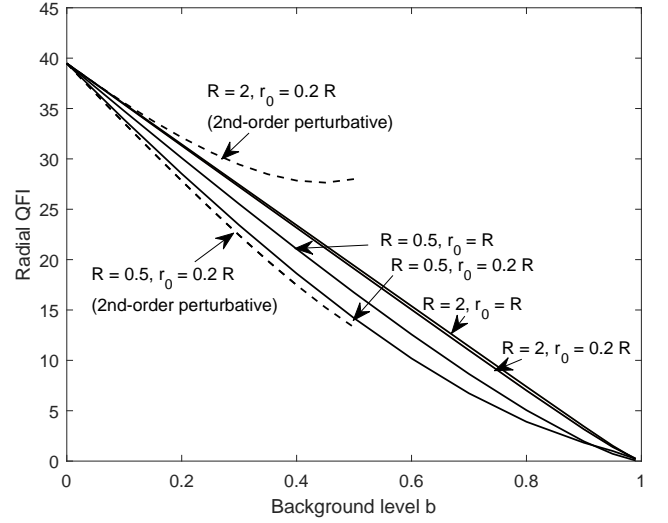}}
\caption{Plot of QFI for estimating the radial distance of the point source from the background disk center vs. the background brightness parameter $b$. The solid curves are the numerically exact results obtained by evaluating Eq.~(\ref{Hmunu2}), while the dashed curves show the corresponding  numerically calculated values of the second-order perturbative expression (\ref{QFI2orders2}).   }
\end{figure}

Figure 3 displays the variation of the azimuthal-localization QFI, $H_{22}$. Its reciprocal is  the minimum possible variance of unbiased estimation of the azimuthal coordinate, which we define as $r_0$ times the angle $\phi_0$ that the source location vector makes with respect to the $x$ axis. As expected for a background disk that is fully rotationally invariant, it is independent of the actual value of $\phi_0$. The trends are largely identical to those seen in Fig.~2 for the QFI for radial-distance estimation, except for the big difference, which we cannot explain, that the second-order perturbative result here remains accurate even out to $b=0.5$ for {\em all} values of $R$ out to 2. 
\begin{figure}
\centerline{
\includegraphics[width=0.55\textwidth]{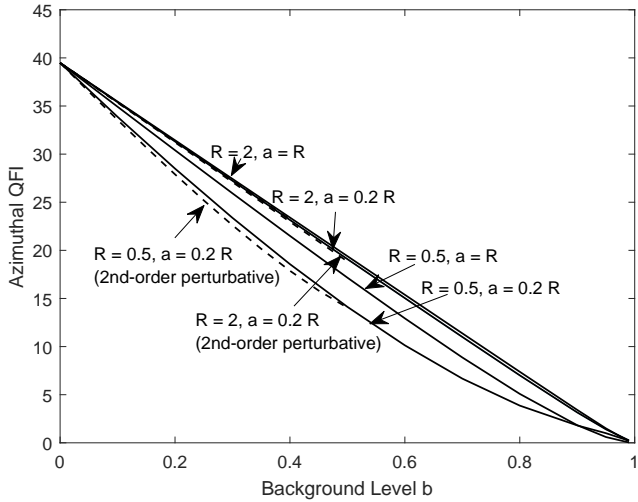}}
\caption{Same as Fig.~2 except that azimuthal-localization QFI is being plotted on the vertical axis.}
\end{figure}

We plot in Figs. 4 and 5 the variation of the radial and azimuthal QFI with respect to the radial distance, $r_0$, of the point source for a moderately strong background at $b=0.5$, with the solid lines showing the numerically exact results and the dashes lines the second-order perturbative results. The discrepancy between the exact and perturbative results for the radial QFI plots shown in Fig.~4 is on the whole larger, the larger $R$ is, reflecting a behavior we have already noted in the plots of Fig.~2 for $R=2$. By contrast, this discrepancy is quite small for the azimuthal QFI plots shown in Fig.~5, mirroring the excellent agreement between the two results even for moderately large values of $b$ that we saw in Fig.~3. We also note that the exact value of neither the radial nor the azimuthal QFI varies dramatically with varying $r_0/R$, particularly for the two larger values of $R$ shown here. The sharper increase of the QFI with increasing $r_0/R$ for the case of a sub-diffractive background disk radius, $R=0.5$, is physically sensible, since at such sub-diffractive scales locating a point source at even smaller distances from the disk center is expected to entail larger localization error when the source is closer to the disk center rather than near the disk boundary.  
\begin{figure}
\centerline{
\includegraphics[width=0.55\textwidth]{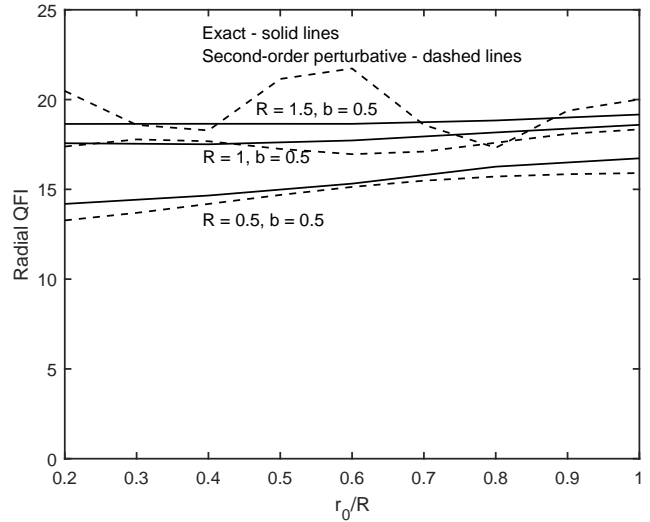}}
\caption{Plot of QFI for estimating $r_0$ vs. $r_0/R$ for a moderately large value of the background brightness parameter, $b=0.5$, and three different values of $R$, 0.5, 1, and 1.5. The solid curves are the numerically exact results obtained by evaluating Eq.~(\ref{Hmunu2}), while the dashed curves show the corresponding  numerically calculated values of the second-order perturbative expression (\ref{QFI2orders2}). }
\end{figure}

\begin{figure}
\centerline{
\includegraphics[width=0.55\textwidth]{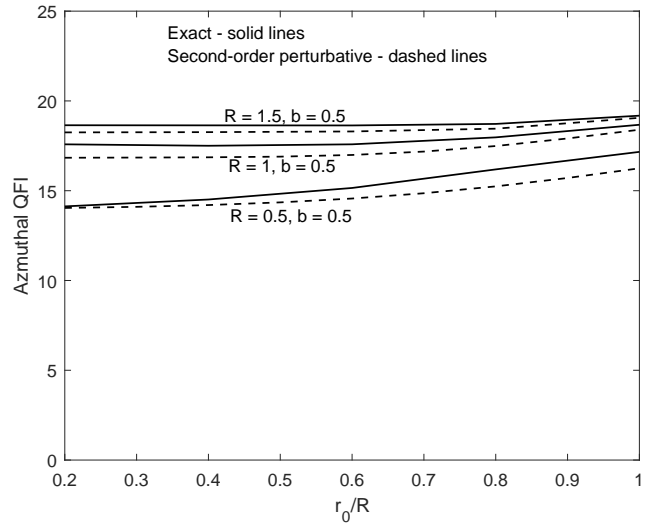}}
\caption{Same as Fig.~4 except that azimuthal-localization QFI is being plotted on the vertical axis.   }
\end{figure}

\section{Localization of a Tiny Hole in a Uniformly Bright Disk}
Consider next photon emission from a uniformly luminous disk of radius $R$ from which a tiny hole of radius $\delta_0 \ll  R$ has been punched out. The problem is one of estimating the position vector $\br_0$ of the hole center relative to the disk center. The geometry for this problem is the same as for the previous problem shown in Fig.~1, in which the point source indicated by a black dot is to be regarded as the brightness hole for the current problem. Its SPDO may be expressed as
\be
\label{H_SPDO}
 \varrho={1\over \pi(R^2-\delta_0^2)}\int_{B_H}dA \ketbra{K_\br}{K_\br},
\ee
where the subscript $B_H$ denotes integration over the disk with the hole. By writing the integral over $B_H$ as the difference between the integral over the full disk, $B$, without holes, and the integral over the hole, $H$, we may express the SPDO (\ref{H_SPDO}) as
\be 
\label{H_SPDO1}
 \varrho =\frac{1}{1-\epsilon}\left(\varrho_B-\epsilon\varrho_H\right),
 \ee
in which $\varrho_B$ is the SPDO corresponding to the uniform disk, as given in Eq.~(\ref{SPDOsb}), and 
\be
\label{H_SPDOh}
\varrho_H\approx \ketbra{K_0}{K_0}
 \ee 
is the SPDO for emission from the hole, were it filled with material of the same irradiance as the rest of the luminous disk, and
 \be
 \label{eps}
 \epsilon={\delta_0^2\over R^2}
 \ee
denotes the ratio of the hole and disk areas, assumed to be small compared to 1. Approximation (\ref{H_SPDOh}), which entails replacing $\int_H dA\ketbra{K_\br}{K_\br}$ by the area of the hole, $\pi\delta_0^2$, times the SPDO corresponding to emission from the center of the hole at $\br_0$, is justified as long as we are not interested in resolving the small hole or estimating its radius, $\delta_0$. Rather, here we only care to estimate the location of its center, $\br_0$, and will thus regard a photon leaving from {\em any} point of the hole in effect as leaving from its center. 

Note the trivial isomorphism between the hole-localization problem and the previously discussed point-source localization problem under the mapping, $b\to (1-\epsilon)^{-1}$. Note also that by replacing the hole-to-disk area ratio $\epsilon$ by $\beta\epsilon$, where $\beta\leq 1$, we may adapt the SPDO model given by Eq.~(\ref{H_SPDO1}) to describe a more general problem in which the radiance in the region of the hole is not zero but a factor $(1-\beta)$ as large as that of the rest of the uniformly bright disk. Without any loss of generality, we will henceforth subsume $\beta$ into the definition of $\epsilon$.

For many applications of interest, such as EP detection \cite{Charbonneau00,Fischer14} for which $\epsilon$ varies between $10^{-3}$ (for Earth-size planets) and $10^{-2}$ (for Jupiter-size planets), we can safely assume that $\epsilon \ll 1$, which justifies the approximation (\ref{H_SPDOh}), and a perturbative treatment of the problem to the lowest order in $\epsilon$ will suffice. Let us expand the SLD in powers of $\epsilon$ as
\be
\label{H_SLDpert}
\cL_\mu=\cL_\mu^{(0)}-\epsilon\cL^{(1)}_\mu+\epsilon^2\cL^{(2)}_\mu+\ldots,
\ee
substitute this expansion and the SPDO (\ref{H_SPDO1}) into Eq.~(\ref{SLDdef}) that defines the SLD to yield the identity,
\be
\label{H_SLDdef}
-\epsilon\pmu \varrho_H={1\over 2}\left[(\cL_\mu^{(0)}-\epsilon\cL^{(1)}_\mu+\ldots)(\varrho_B-\epsilon\varrho_H)+\ha\right],
\ee
where the left hand side contains only the SPDO of the hole whose location we are trying to estimate. Equating the two sides of Eq.~(\ref{H_SLDdef}) power by power in $\epsilon$ yields
\ba
\label{H_SLDrels}
0=&\cL_\mu^{(0)}\varrho_B+\varrho_B\cL_\mu^{(0)};\nn
2\pmu\varrho_H=&\left(\varrho_B\cL_\mu^{(1)}+\varrho_H\cL_\mu^{(0)}\right)+\left(\cL_\mu^{(1)}\varrho_B+\cL_\mu^{(0)}\varrho_H\right);\nn
\vdots \qquad&\qquad\qquad \vdots \hspace{100pt}\vdots
\end{align}

It follows from the first of the relations in Eq.~(\ref{H_SLDrels}) that $\cL^{(0)}_\mu$ must vanish identically. This implies the following form for the QFI matrix elements:
\ba
\label{H_QFIpert}
H_{\mu\nu}=&\Tr\left(\cL_\mu\pnu\varrho\right)\nn
                =&{\epsilon^2\over1-\epsilon}\Tr\left(\cL^{(1)}_\mu\pnu\varrho_H+\order{\epsilon}\right),
\end{align}
in which we used once again the fact that only the $\varrho_H$ part of the SPDO (\ref{H_SPDO}) contributes to the derivative $\pnu\varrho$. We will include and evaluate only the first term on the right-hand side of Eq.~(\ref{H_QFIpert}), since we have assumed that $\epsilon\ll 1$ and thus excellent accuracy is assured despite an omission of all higher order terms. In view of expression (\ref{H_SPDOh}) for $\varrho_H$, we may evaluate the trace in expression (\ref{H_QFIpert}) as
\ba
\label{H_QFIpert2}
H_{\mu\nu}=&\epsilon^2\left(\pnu\bra{K_0}\cL^{(1)}_\mu\ket{K_0}+\bra{K_0}\cL_\mu^{(1)}\pnu\ket{K_0}\right)\nn
                  =&2\epsilon^2\Re \left(\pnu\bra{K_0}\cL^{(1)}_\mu\ket{K_0}\right),
\end{align}
in which in consistency with the neglect of the higher order terms we have neglected $\epsilon$ from the denominator of the overall coefficient.  

To evaluate the RHS of Eq.~(\ref{H_QFIpert2}), we return to the second of the relations in Eq.~(\ref{H_SLDrels}), with $\cL^{(0)}_\mu$ set to 0, and then transform it into the Lyapunov solution for $\cL^{(1)}_\mu$ using the approach used to derive Eq.~(\ref{Lyapunov}),
\be
\label{H_Lyapunov}
\cL^{(1)}_\mu=2\lim_{\eta\to 0^+}\int_0^\infty dx \exp(-x\varrho_\eta)\pmu\varrho_H\exp(-x\varrho_\eta),
\ee
in which $\varrho_\eta$ is defined as a full-rank extension of the background-disk SPDO, analogous to that in Eq.~(\ref{FullRankDO}),
\be
\label{H_FullRankDO}
\varrho_\eta=(1-\eta)\varrho_B+{\eta\over D}\cI,
\ee
with $D$ defined as before. In terms of the sets of eigenvalues, $\{\lambda_i\mid i=1,2,\ldots\}$, and corresponding orthonormal eigenstates, $\{\ket{\lambda_i}\mid i=1,2,\ldots\}$, of $\varrho_B$, we may express the exponential operator in Eq.~(\ref{H_Lyapunov}) as
\be
\label{H_exp}
\exp(-x\varrho_\eta)=\exp(-x\eta/D)\sum_i\exp[-x(1-\eta)\lambda_i]|\ketbra{\lambda_i}{\lambda_i},
\ee
and thus the state resulting from the action of $\cL^{(1)}_\mu$  on $\ket{K_0}$ as the double sum
\ba
\label{H_SLDmuK0}
\cL^{(1)}_\mu\ket{K_0}=&2\sum_{i,j}\ket{\lambda_i}{\mel{\lambda_i}{\pmu\varrho_H}{\lambda_j}\over \lambda_i+\lambda_j}\braket{\lambda_j}{K_0},
\end{align}
in which it is understood that $\lambda_i+\lambda_j\neq 0$ for all the terms that are included in the double sum. This latter restriction is necessary in order to apply the limit, $\eta\to 0^+$, after performing the integral over $x$ in Eq.~(\ref{H_Lyapunov}). 

By taking the inner product of expression (\ref{H_SLDmuK0}) with $\pnu\bra{K_0}$ and adding to the resulting inner product its complex conjugate, we obtain from Eq.~(\ref{H_QFIpert2}), as shown in detail in Appendix C, the following expression for the QFI matrix element:
\ba
\label{H_QFIpert3}
H_{\mu\nu}=&2\epsilon\,\pnu\!\bra{K_0}\cI_N\pmu\ket{K_0}\nn
                  +&2\epsilon^2\Big[\sum_{i,j\in\cS}{\lambda_i\lambda_jC_i(\br_0)C_j(\br_0)\pnu\!\braket{K_0}{\lambda_i}\pmu\!\braket{K_0}{\lambda_j}\over\lambda_i+\lambda_j}\nn
+&\sum_{i,j\in\cS}{\lambda_j^2C_j^2(\br_0)\pnu\!\braket{K_0}{\lambda_i}\bra{\lambda_i}\pmu\!\ket{K_0}\over\lambda_i+\lambda_j}\Big].
\end{align}       
%in which $\cS$ and $\cN$ are the sets of indices labeling the eigenstates of the background-disk SPDO in its support and null subspaces, respectively. 
The eigenfunctions $C_i(\br)$, obeying the following integral equation over the disk,
\be
\label{IntegralEqn}
{1\over \pi R^2}\int_B dA'  {2J_1(2\pi|\br-\brp|)\over 2\pi |\br-\brp|} C_i(\brp) =\lambda_i C_i(\br),
\ee
have all been chosen to be real. The matrix elements that need to be evaluated in expression (\ref{H_QFIpert3}) are all of the same type, namely $\bra{\lambda_i}\pmu\ket{K_0}$, which we have already evaluated as the disk-area integral (\ref{lambdai_pmu_K0}) involving the eigenfunction $C_i(\br)$.% and with $b$ set equal to 1.

A detailed analysis of the disk SPDO problem, including the method of numerical computation of its eigenstates and eigenvalues, was presented in Ref.~\cite{Prasad20b}. We use the results of that eigenanalysis here to evaluate the QFI matrix elements of Eq.~(\ref{H_QFIpert3}) to complete the perturbative treatment of our problem of estimating the location of the hole in the otherwise uniformly luminous, incoherent, disk-shaped background. As we also noted earlier for the problem of locating a point source against the disk background, the QFI matrix once again becomes diagonal when referred to Cartesian coordinate axes that are parallel and perpendicular to the radial location vector of the brightness hole.
 
It is interesting to note that since $\epsilon \ll 1$, the first term on the RHS of expression (\ref{H_QFIpert3}), which is seemingly proportional to $\epsilon$, might dominate the double-sum terms that are proportional to $\epsilon^2$. In our numerical evaluations, however, the opposite turns out to be true, with the first term being negligibly small or identically zero to the numerical accuracy of our computations. This seems to support our conjecture that the SPDO for the uniformly bright disk might be nearly full-rank, if not exactly so. 

\begin{figure}
\centerline{
\includegraphics[width=0.55\textwidth]{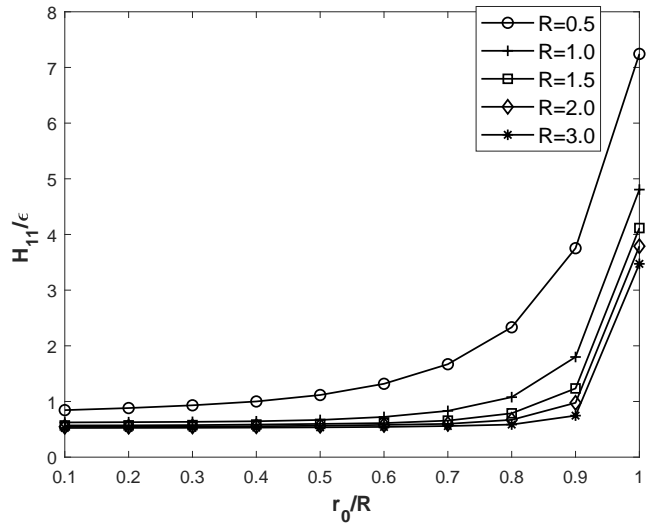}}
\caption{Plots of $H_{11}/\epsilon$ for estimating the radial coordinate of the hole from the background disk center vs. the radial distance of the hole, $r_0$, for five different values of the disk radius. }
\end{figure}

\begin{figure}
\centerline{
\includegraphics[width=0.55\textwidth]{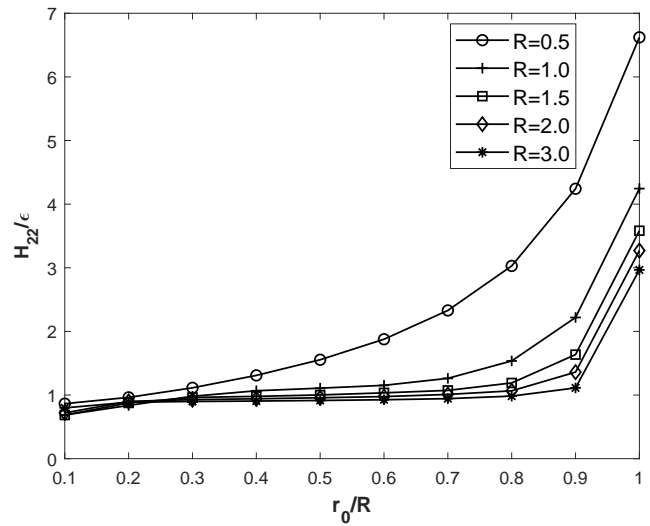}}
\caption{Plots of $H_{22}/\epsilon$ for estimating the azimuthal coordinate of the hole from the background disk center vs. $r_0$ for the same values of $R$ as in Fig.~6.}
\end{figure}

In Figs.~6 and 7, we plot the radial and azimuthal QFI, given by $H_{11}$ and $H_{22}$, respectively, vs. the fractional radial distance, $r_0/R$, of the brightness hole from the center of the background disk. We have scaled the vertical axis by $\epsilon$, the ratio of the hole to disk areas, to show that the QFI for both radial and azimuthal hole-position estimations is essentially inversely proportional to the disk area, $\pi R^2$ for values of $R$ greater than 1, as we clearly see from the closely spaced curves corresponding to $R=1.5$, $R=2$ and $R,=3$ in both figures. Equivalently, the quantum Cram\'er-Rao bounds, given by the reciprocals $1/H_{11}$ and $1/H_{22}$, scale proportional to the ratio of the disk area to the hole area. A heuristic understanding of this fact can be gained by recognizing that the problem is one of determining which specific one of the $N=1/\epsilon$ contiguously filling segments of the disk, with each segment being of area equal to the hole area, contains the actual brightness hole. The variance of such determination must scale proportional to that number when the disk brightness is otherwise uniform everywhere and there is no preferred location of the hole. Equivalently, only the photon leaving from the hole carries any information about its location, so the information fundamentally must be proportional to the probability of such event, which for an otherwise uniformly bright disk is simply the ratio of the hole area to the disk area, $\epsilon$. In the plots, we fixed the radius of the hole at 0.05, so for the five values of the disk radius shown in these figures, the number $1/\epsilon$ varies between 100 (for $R=0.5$) and 3600 (for $R=3$). 

A second observation relates to the increase of the QFI for both radial and azimuthal estimation with increasing fractional radial distance of the hole from the disk center, $r_0/R$. The increase is particularly pronounced for the smallest disk radius, $R=0.5$, which is in the sub-diffractive regime. This is due to the fact that as the hole gets farther out from the center of a sub-diffractive-scale disk, one expects to optically locate the position of the hole with greater statistical confidence despite diffraction-induced positional uncertainties. For the larger disk radii too, the QFI for estimating the hole location relative to the disk center increases with increasing radial distance of the hole, particularly as the hole gets closer to the edge of the disk. For a uniformly bright disk, the probability of information-bearing photons leaving, for example, from its outer half area in the radial range, $(R/\sqrt{2},R)$, is the same as that for photons leaving from its inner half area in the radial range, $(0,R/\sqrt{2})$, with the two ranges having radial widths in the ratio of 0.29/0.71, which is considerably smaller than 1. As a result, the hole if located in the outer annular half of the disk area has a smaller radial position uncertainty than if it is located in the inner half. This expectation is confirmed by the monotonic increase of the QFI with the radial distance of the hole from the disk center.  

\section{Wavefront Projections and Source/Hole Localization}

We next consider the use of projection of the imaging wavefront into a set of orthonormal modes \cite{Tsang16} as a way of recovering information at the quantum level about the 2D location of the point source and of the hole against the disk-shaped uniform-brightness background. We first evaluate the Cram\'er-Rao lower bounds (CRBs) for the source localization problem for three different sets of modes, namely the Zernike (Z) \cite{Noll76}, Fourier-Bessel (FB) \cite{Watson95,Lebedev72}, and localized-source modes. We have already demonstrated the efficacy of the Zernike modes for estimating the location and separation of a pair of point sources \cite{YuPrasad18,PrasadYu19}, under finite-bandwidth emission \cite{Prasad20c}, and for estimating the physical size parameters of extended sources in one and two dimensions \cite{Prasad20b}, including the radius of a uniformly bright disk. %We will borrow many mathematical expressions needed for our present analysis from Ref.~\cite{Prasad20b}. 

The FB modes furnish an alternate set of modes which, like the Zernikes, have been used in a variety of conventional optics applications, e.g., for representing the phase of a turbulence-degraded wavefront in astronomy \cite{Hart05}, for principal-component analysis of images \cite{Zhao13}, and for efficient numerical electromagnetics \cite{Dems21}. The final basis that we consider here consists of a set of orthogonalized modes constructed from planar wavefunctions emitted by a set of 30-40 localized fictitious point sources distributed over the background disk. Under a variety of operating conditions, the last set, despite its finite cardinality, seems to perform the best of the three in terms of how closely its CRB approaches the corresponding QCRB bound for both localization problems.
% but they seem to be the least effective basis in localizing the point source of interest, particularly in the super-diffractive regime of large disk radius, as we will see presently.

We may regard conventional clear-aperture imaging as a variety of wavefront projection into contiguous non-orthogonal modes, namely the complex Fourier modes representing a continuously varying wavefront tilt over a finite aperture. We can therefore disuss direct imaging in the present section itself.

\subsection{Direct Imaging}
We may regard the point-spread function (PSF), when normalized to have unit area, in a conventional imaging protocol as the result of projection of a clear (or suitably coded or apodized) wavefront emitted by a single point source into contiguous complex exponential Fourier modes that yield the image at different pixels of the imaging sensor. For a finite aperture, these Fourier modes are nonorthogonal, but the corresponding projections onto the sensor pixels yield image intensities that represent, in a mutually statistically independent fashion, the probabilities of an imaging photon to land at different sensor pixels. When the photons are emitted by a more complex scene, such as the ones of interest here, namely a point source or a brightness hole in an otherwise uniformly bright background, the probabilities of each photon landing at different sensor pixels are modified by the intensity distribution of the scene. At the single photon level, the direct-imaging (DI) probability of the photon landing at a unit area centered at the image point $\br$ is defined as the expectation value,
\be
\label{DIprob}
P(\br)=\pi \mel{K_{\br}}{\varrho}{K_{\br}}.
\ee
We will assume here that the aperture indicator function $\Theta_P(\bu)$ in terms of which the aperture-plane representation of the projection modes $\ket{K_{\br}}$ is defined by Eq.~(\ref{wavefunction}) is that of an uncoded clear aperture, but more complicated coded apertures may be used if necessary for specific applications.

With the probability density (\ref{DIprob}), which integrates to 1 over the unbounded image plane, we may calculate the $\mu\nu$ element of the FI matrix per photon for estimating the location of the point source against the background disk in the photon-counting limit as the area integral \cite{Tsang16},
\be
\label{DI_FI}
I_{\mu\nu}=\int {\pmu P(\br)\pnu P(\br)\over P(\br)} dA,
\ee 
over the full image plane. Note that the probability density $P(\br)$ also represents the mean image intensity per photon, which may be expressed in terms of the PSF, $h(\br)$, as 
\be
\label{DIint}
P(\br) = (1-b) h(\br-\br_0)+{b\over \pi R^2}\int_B dA' h(\br-\br').
\ee
For a circular clear aperture, the unit-area PSF has the Bessel form,
\be
\label{DI_PSF}
h(\br-\br') = \pi|\braket{K_{\br}}{K_{\br'}}|^2= {J_1^2(2\pi|\br-\br'|)\over\pi |\br-\br'|^2}. 
\ee
Representing the FI by continuous integrals of form (\ref{DI_FI}) assumes that the image plane is unpixelated and continuous, but in any practical camera the FI will, in general, have a smaller value due to the finite size and unavoidable detection noise of its discrete pixels.  

For localizing a hole in the disk background, our second problem, the probability density, $P(\br)$, is readily obtained from Eq.~(\ref{DIint}) by recognizing the formal difference between expressions (\ref{SPDO1}) and (\ref{H_SPDOh}), entailing merely a replacement of $b$ by $1/(1-\epsilon)$ and $1-b$ by  $-\epsilon/(1-\epsilon)$ in Eq.~(\ref{Zprob}). This being the only difference between the expressions for the single-photon probability density for the two problems, we do not discuss them any further here and in the subsequent subsections on the other wavefront projection bases. In view of such isomorphism between the two problems, we will display detailed results for direct imaging for the first problem alone when comparing with other wavefront projection bases.  
 
\subsection{Zernike modes}
\begin{figure}[htb]
\centering
\subfloat[]
{\includegraphics[width=0.22\columnwidth]{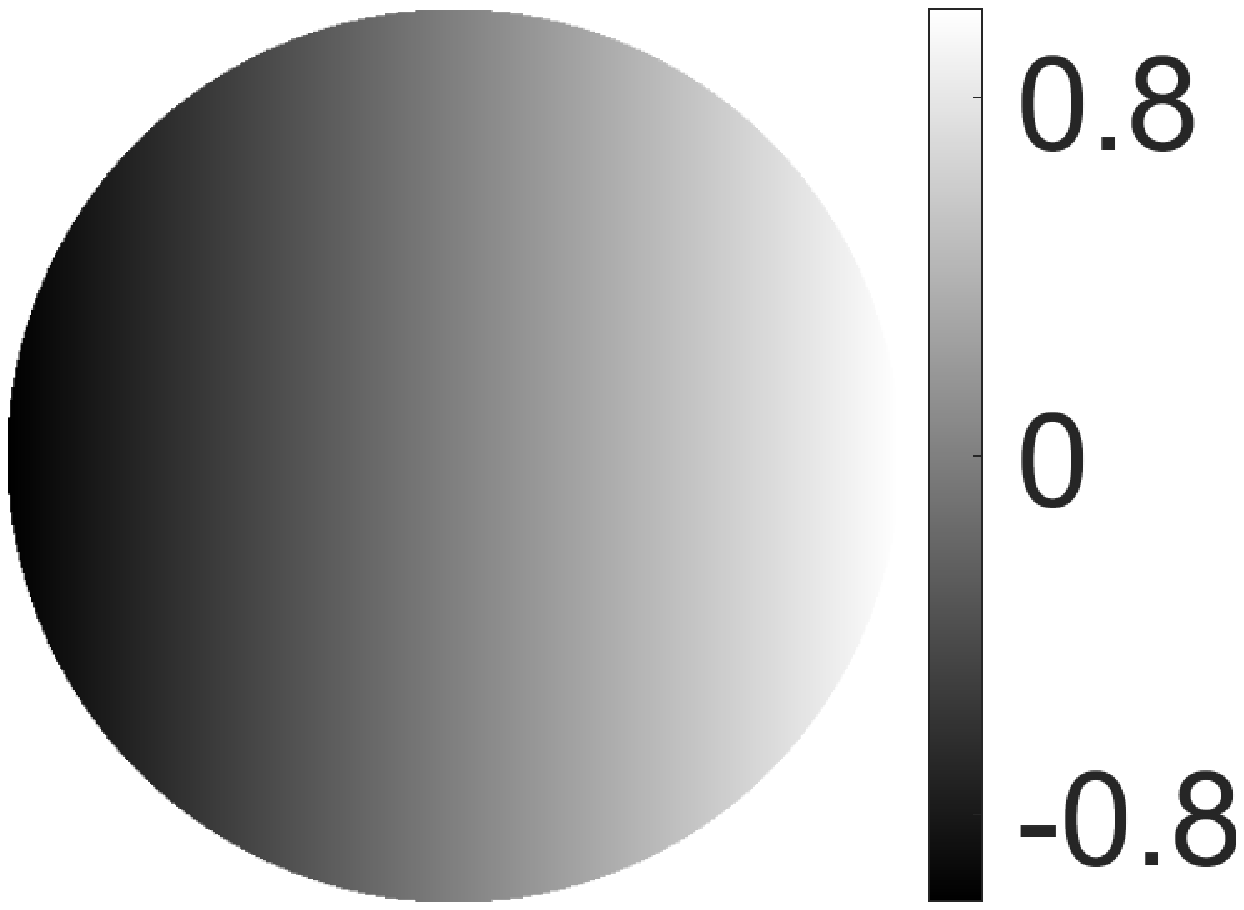}}
\subfloat[]
{\includegraphics[width=0.22\columnwidth]{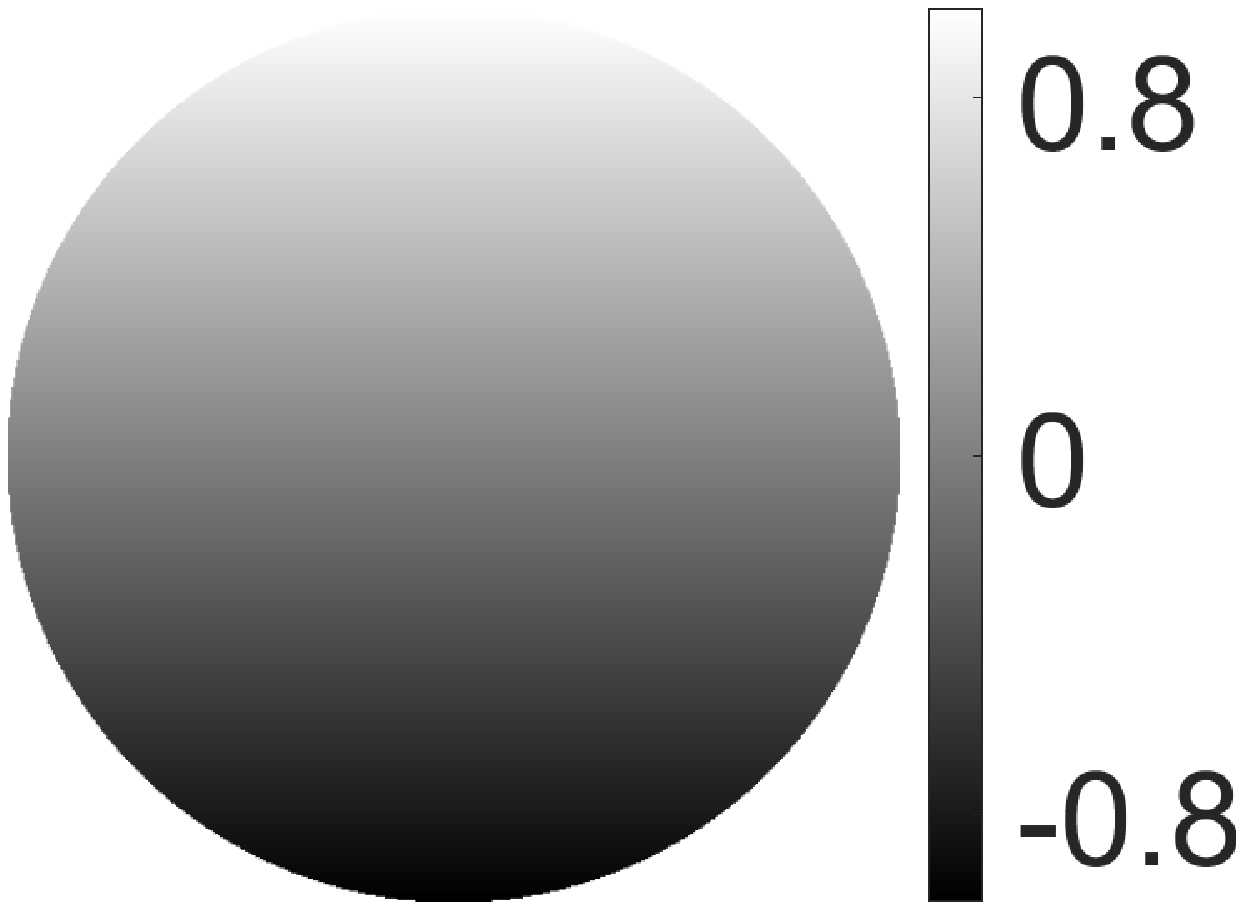}}
\subfloat[]
{\includegraphics[width=0.22\columnwidth]{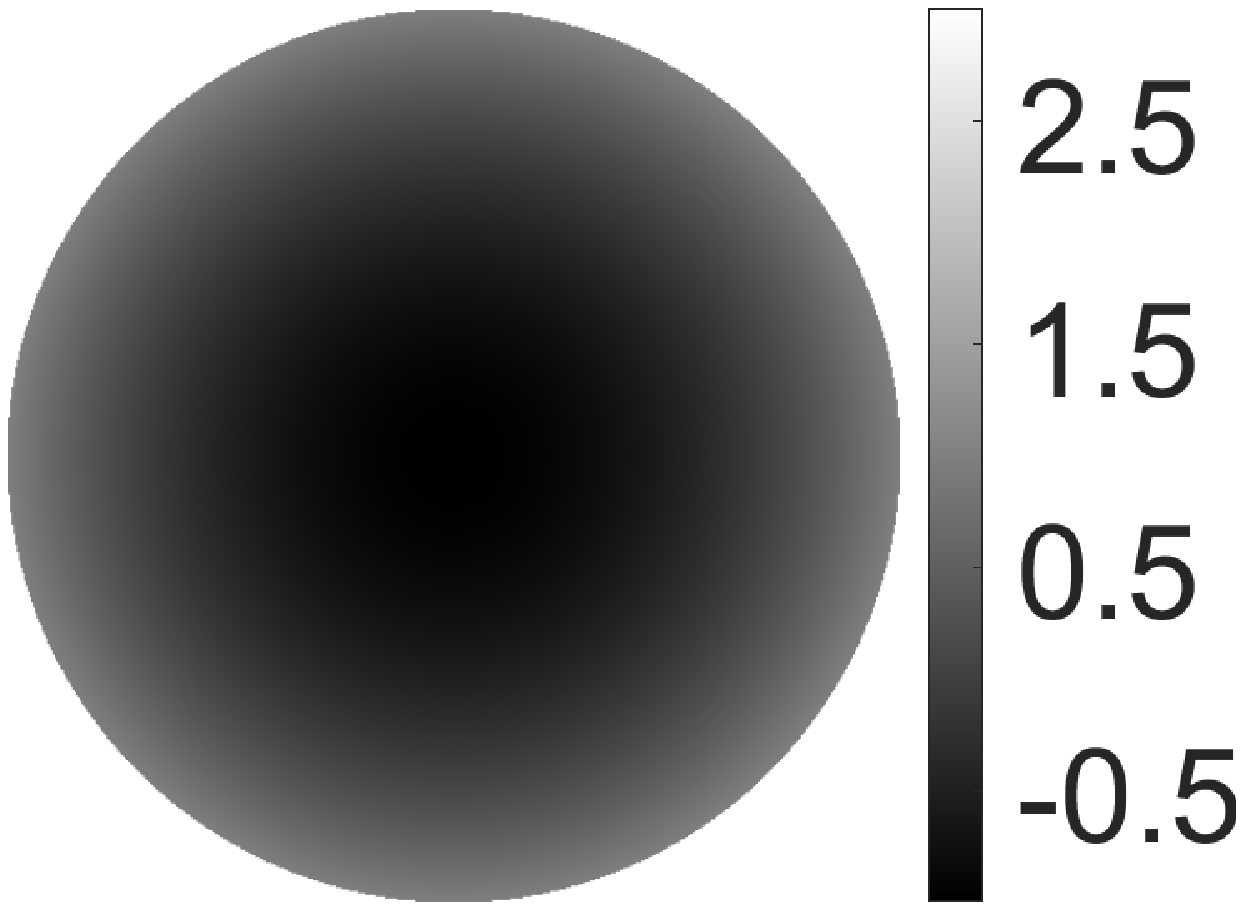}}
\subfloat[]
{\includegraphics[width=0.22\columnwidth]{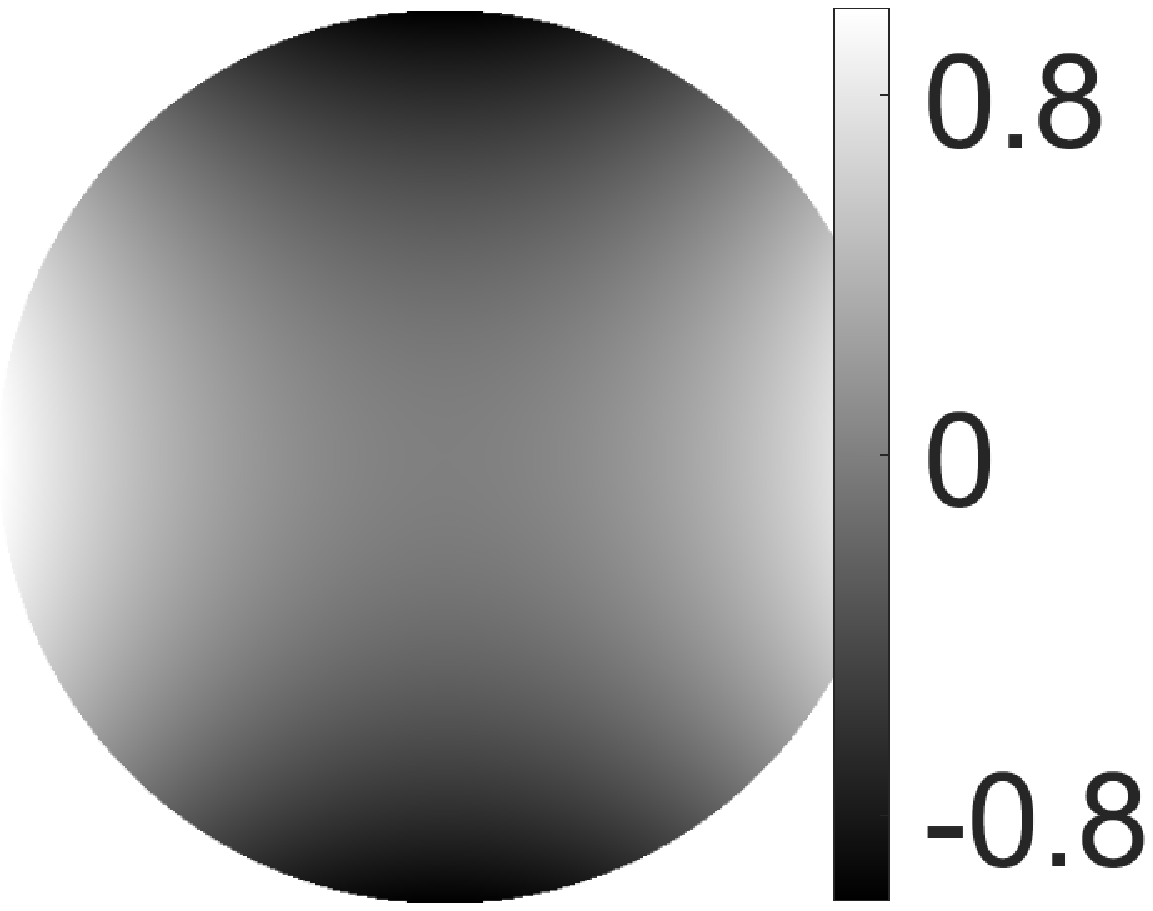}}\\
\caption{\label{fig:Zernikesurface} Surface plots of (a) $Z_{11,+1}$ (b) $Z_{11,-1}$, (c) $Z_{20}$, and (d) $Z_{22,+1}$. Vertical bars at the right of each plot indicate the range of the grayscale.}
\end{figure}

The Zernike modes, defined to be non-vanishing only over the unit disk, $|\bu|\leq 1$, are labeled by two non-negative integer indices, $p$ and $m$, with $p=0,1,2,\ldots$ and $m=p,p-2,\ldots$, and a mirror-reflection-symmetry index $\sigma$, taking the values $\pm1$ corresponding to cosine and sine azimuthal-angular dependences. 
Surface plots of four low-order Zernike modes are given in Fig.~\ref{fig:Zernikesurface}.

Of specific interest here is the 2D Fourier transform (FT) property \cite{Prasad20b} of the Zernike modes, 
\ba
\label{ZFT}
\int &d^2u P(\bu)\exp(-i2\pi \bu\cdot\br) \, Z_{pm\sigma}(\bu) =\sqrt{p+1\over\pi}{J_{p+1}(2\pi r)\over r}\nn
&\times\left\{
\begin{array}{ll}
(-1)^{(p-m)/2}(-i)^m\sqrt{2}\cos m\phi, &  m\neq 0,\, \sigma=+1\\
(-1)^{(p-m)/2}(-i)^m\sqrt{2}\sin m\phi, & m\neq 0,\, \sigma=-1\\
(-1)^{p/2},& m=0.
\end{array}
\right.
\end{align}
In view of the point-source wavefunctions (\ref{wavefunction}), which are simply proportional to the complex Fourier exponentials, the FT property (\ref{ZFT}) immediately yields the projection probabilities of a point-source wavefunction in the Zernike modes to be
\ba
\label{ZprobPtSource}
|\braket{Z_{pm\sigma}}{K_\br}&|^2= {(p+1)\over\pi^2}{J_{p+1}^2(2\pi r)\over r^2}\nn
&\times\left\{
\begin{array}{ll}
2\cos^2 m\phi, & m\neq 0,\, \sigma=+1 \\
2\sin^2 m\phi, & m\neq 0,\, \sigma=-1\\
1,& m=0,
\end{array}
\right.
\end{align}
and thus the projection probabilities of the single-photon density operator (\ref{SPDO1}) in the Zernike modes as
\ba
\label{Zprob}
P_{pm\sigma} =&\mel{Z_{pm\sigma}}{\varrho}{Z_{pm\sigma}}\nn
=&{(1-b)(p+1)\over\pi^2}{J_{p+1}^2(2\pi r_0)\over r_0^2}\nn
&\times\left\{
\begin{array}{ll}
2\cos^2 m\phi_0, & m\neq 0,\, \sigma=+1 \\
2\sin^2 m\phi_0, & m\neq 0,\, \sigma=-1\\
1,& m=0.
\end{array}
\right.\nn
&+{2b(p+1)\over \pi^2 R^2}\int_0^R dr{J_{p+1}^2(2\pi r)\over r}.
\end{align}
To reach the final term in expression (\ref{Zprob}), we performed the integral over the full range $(0,2\pi)$ of the azimuthal angle over the disk, which simply yields $2\pi$ for each of the angular dependences of form (\ref{ZprobPtSource}). The remaining radial integral over the disk in the final term can be evaluated exactly as a sum of squares of Bessel functions of different orders, as shown in Appendix B of Ref.~\cite{Prasad20b}.

The partial derivatives of expression (\ref{Zprob}) with respect to $r_0$ and $\phi_0$, the polar coordinates of the position of the point source being estimated, involve only its first term, and are easily calculated analytically. With expression (\ref{Zprob}) and its partial derivatives in hand, we can now calculate the FI for simultaneously estimating the radial distance, $r_0$, and its azimuthal, arc-length coordinate, $r_0\phi_0$, as the matrix
\be
\label{FIrphi}
\bI=
\begin{pmatrix}
I_{r_0r_0} &I_{r_0\phi_0}/r_0\\
I_{r_0\phi_0}/r_0 & I_{\phi_0\phi_0}/r_0^2,
\end{pmatrix}
\ee 
in which each matrix element is defined by the following sum over the modes, labeled generically by a single index $\lambda$:
\be
\label{FIMatEl}
I_{\mu\nu}=\sum_{\lambda}{\partial_\mu P_\lambda\,\pnu P_\lambda\over P_\lambda}.
\ee
 
\subsection{Fourier-Bessel modes}
\begin{figure}[htb]
\centering
\subfloat[]
{\includegraphics[width=0.22\columnwidth]{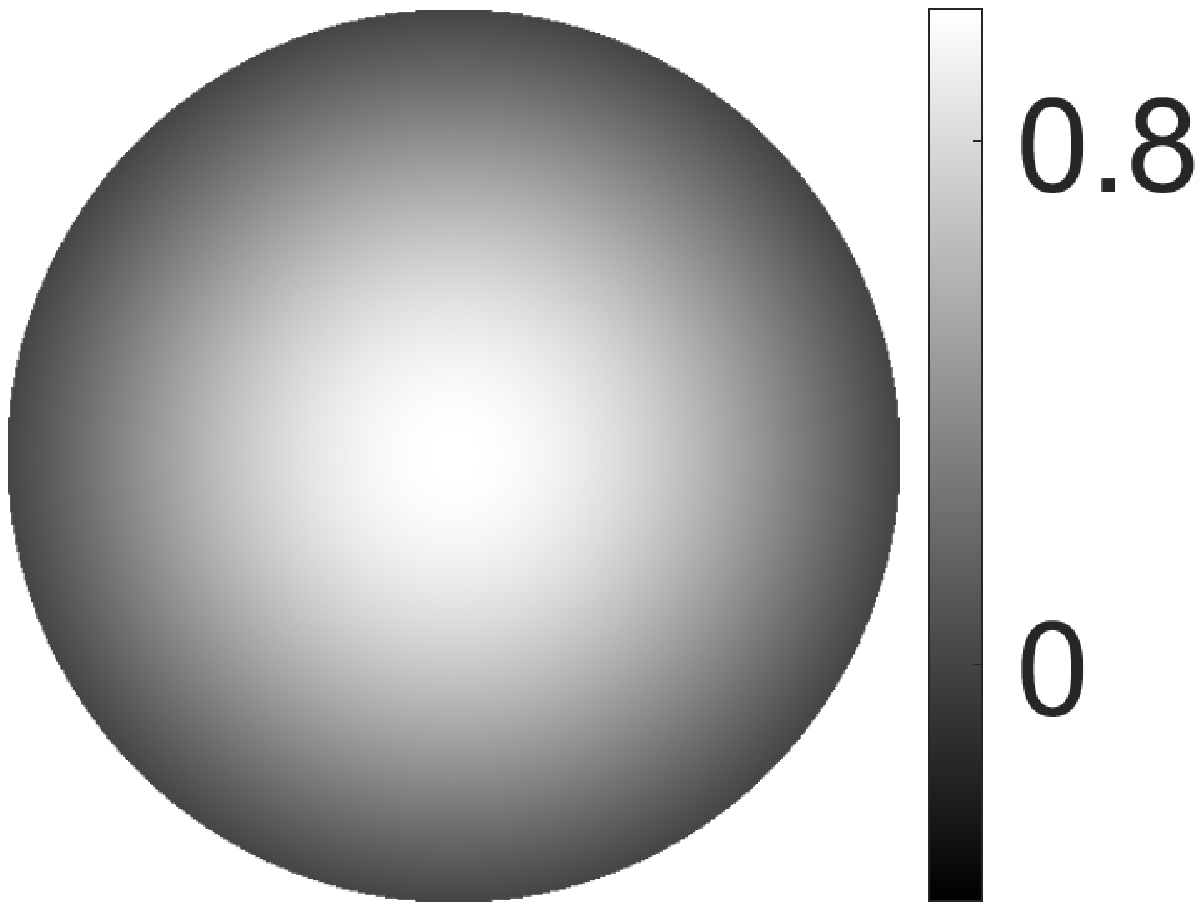}}
\subfloat[]
{\includegraphics[width=0.22\columnwidth]{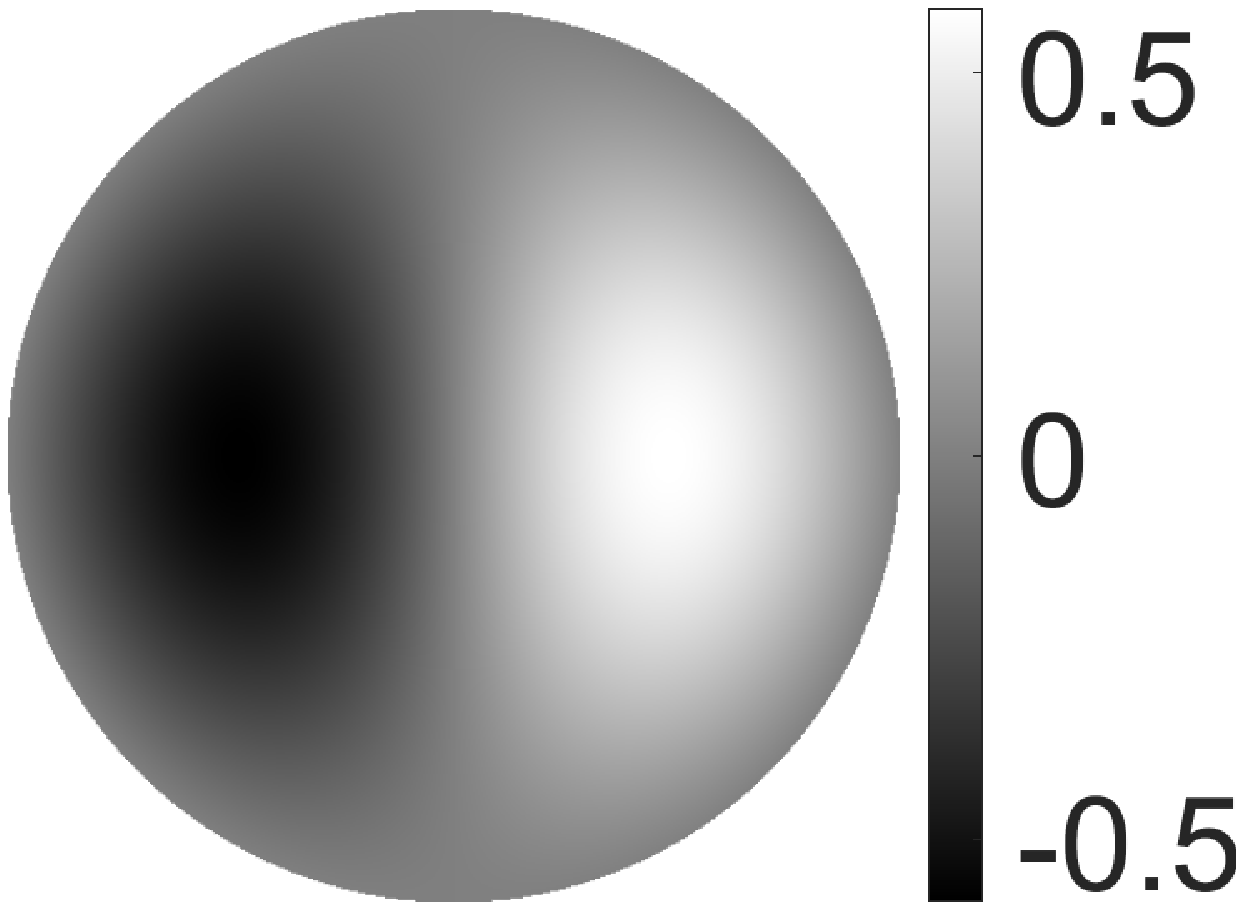}}
\subfloat[]
{\includegraphics[width=0.22\columnwidth]{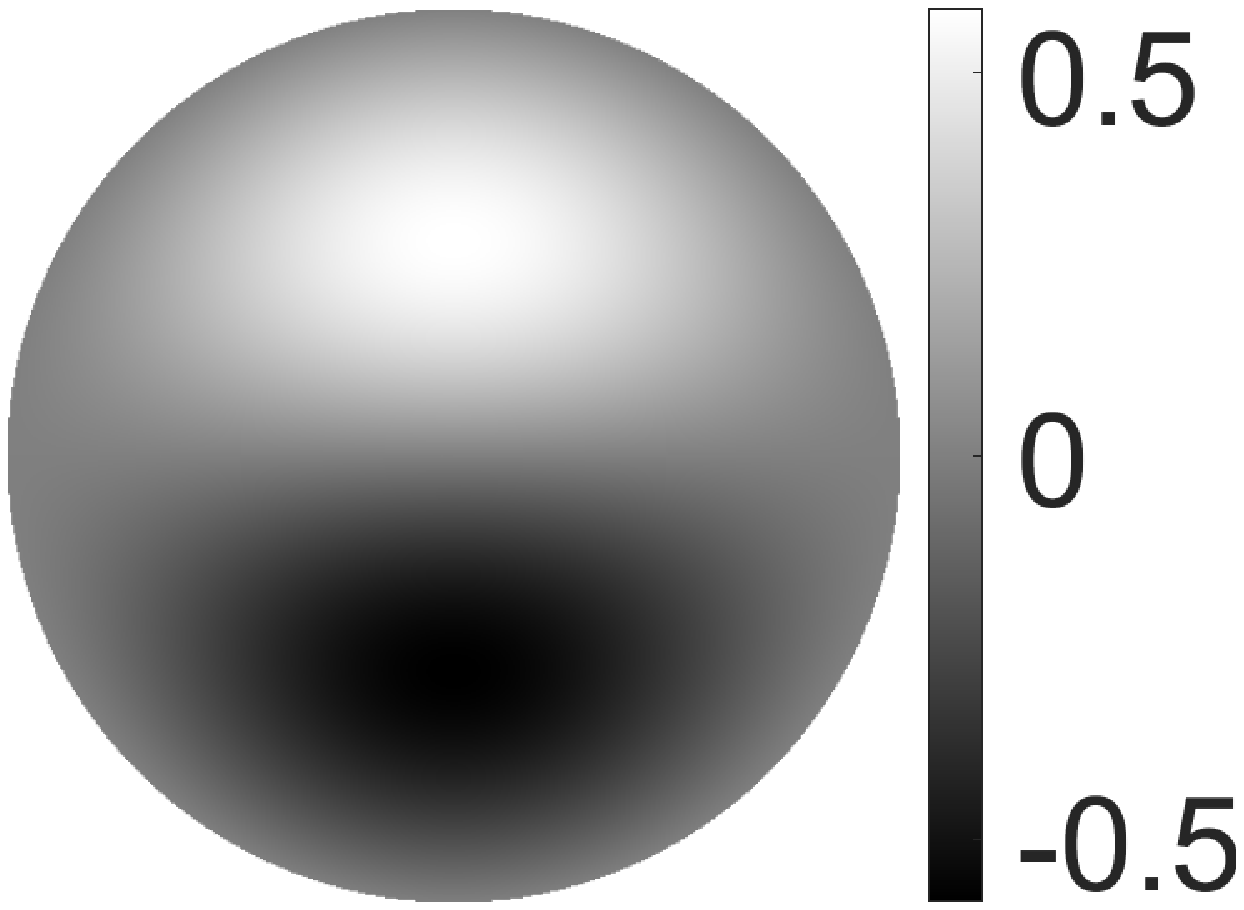}}
\subfloat[]
{\includegraphics[width=0.22\columnwidth]{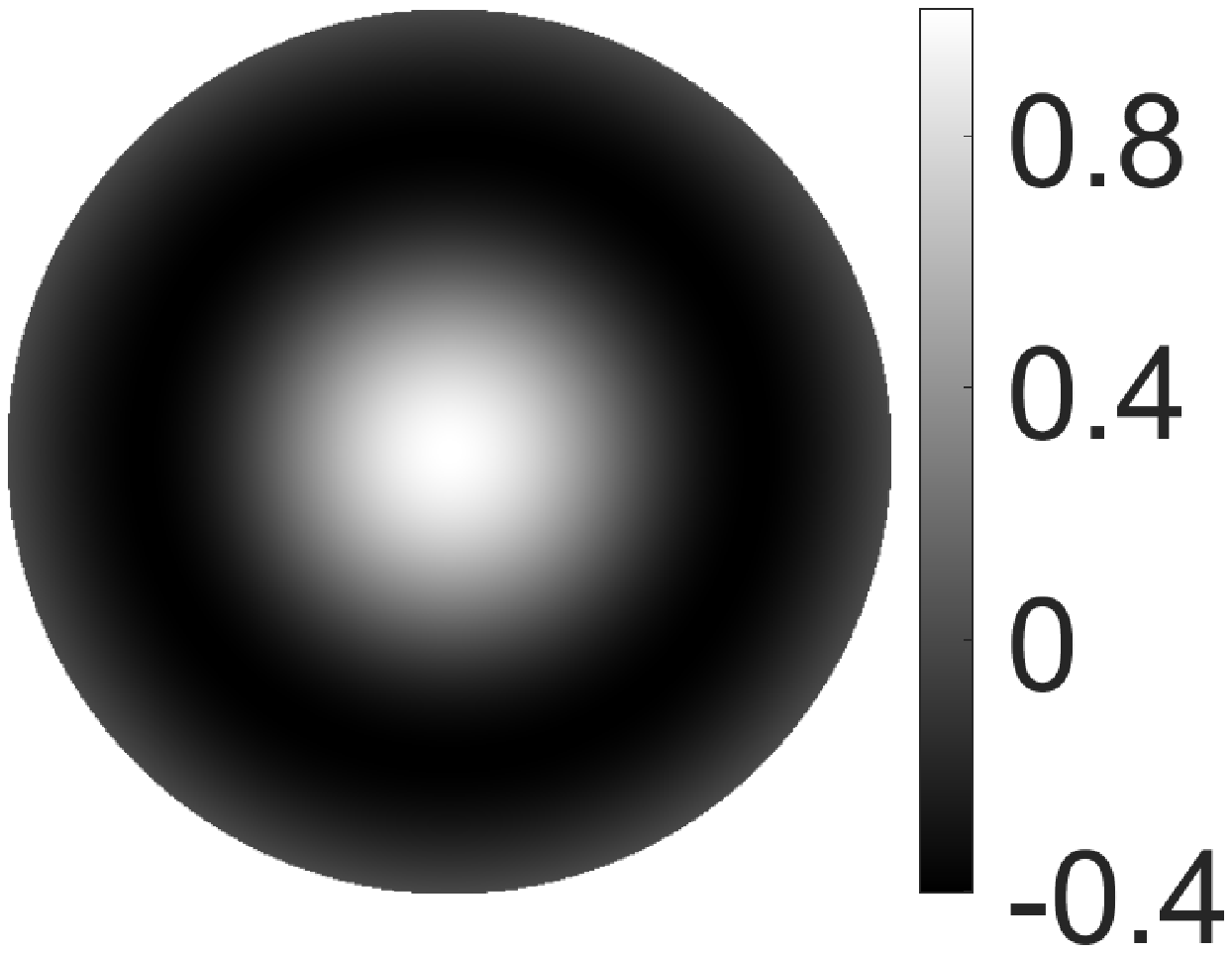}}\\
\caption{\label{fig:FBsurface} Surface plots of (a) $F_{01}$ (b) $F_{11,+1}$, (c) $F_{11,-1}$, and (d) $F_{02}$. Vertical bars at the right of each plot indicate the range of the grayscale. Each function vanishes at the edge of the circular aperture.}
\end{figure}

The Fourier-Bessel (FB) modes, sometimes called disk harmonics, are defined in their real version as the complete set of orthonormal functions over the unit disk $0\leq u\leq 1,\ 0\leq \phi_u < 2\pi$ of the form \cite{Lebedev72},
\ba
\label{FB}
F_{mn\sigma}(\bu) &={J_{p}(x_{mn}u)\over \sqrt{\pi}J_{m+1}(x_{mn})}\nn
&\times\left\{
\begin{array}{ll}
\sqrt{2}\cos m\phi_u, &  m\neq 0,\, \sigma=+1\\
\sqrt{2}\sin m\phi_u, & m\neq 0,\, \sigma=-1\\
1,& m=0,
\end{array}
\right.
\end{align}
in which $x_{mn}$ is the $n$th positive zero of Bessel function $J_m(\cdot)$ and the mode indices take the values, $m=0,1,\ldots$, $n=1,2,\ldots$, and $\sigma=\pm 1$. Like the Zernikes, they are normalized over the unit disk,
\ba
\label{FBnorm}
\braket{F_{mn\sigma}}{F_{m'n'\sigma'}}=&\int dA \, P(\bu)\, F_{mn\sigma}(\bu)\,F_{m'n'\sigma'}(\bu)\nn
                                                            =&\delta_{mm'}\delta_{nn'}\delta_{\sigma\sigma'}.
\end{align}
Surface plots of four low-order FB modes are given in Fig.~\ref{fig:FBsurface}.

The 2D FT of these functions yields their inner product with the single-photon wavefunction (\ref{wavefunction}) emitted by a point source and transmitted to the clear circular aperture of the imaging system,
\ba
\label{FBKr}
\braket{F_{mn\sigma}}{K_\br}=&{1\over\pi J_{m+1}(x_{mn})}\int d^2u\, P(\bu)\, \exp(-i2\pi\bu\cdot\br) \nn
             &\times J_m(x_{mn}u)\left\{
\begin{array}{ll}
\sqrt{2}\cos m\phi_u, &  m\neq 0,\, \sigma=+1\\
\sqrt{2}\sin m\phi_u, & m\neq 0,\, \sigma=-1\\
1,& m=0
\end{array}
\right.\nn
=&{2(-i)^m\over J_{m+1}(x_{mn})}\int_0^1 du\, u\, J_m(2\pi ru)\, J_m(x_{mn}u)\nn
             &\times\left\{
\begin{array}{ll}
\sqrt{2}\cos m\phi, &  m\neq 0,\, \sigma=+1\\
\sqrt{2}\sin m\phi, & m\neq 0,\, \sigma=-1\\
1,& m=0,
\end{array}
\right.\nn
\end{align}
in which we used the following Bessel integral identity:
\ba
\label{BesselIdentity}
\int_0^{2\pi} d\phi_u &\exp[-iz\cos(\phi-\phi_u)]\,\left\{\begin{array}{l}\cos m\phi_u\\ \sin m\phi_u\end{array}\right.\nn
=&2\pi(-i)^mJ_m(z)\left\{\begin{array}{l}\cos m\phi\\ \sin m\phi\end{array}\right.
\end{align}
The radial ($u$) integral in Eq.~(\ref{FBKr}) may be evaluated in closed form \cite{Lebedev72} as
\ba
\label{RadInt}
\int_0^1 du\, u\, &J_m(2\pi ru)\, J_m(x_{mn}u)\nn
=&{2\pi rJ_m(x_{mn})J_m^\prime(2\pi r)-x_{mn}J_m^\prime(x_{mn})\,J_m(2\pi r)\over x_{mn}^2-(2\pi r)^2}\nn
=&{x_{mn} J_m(2\pi r)\, J_{m+1}(x_{mn})\over x_{mn}^2-(2\pi r)^2},
\end{align}
in which the prime superscript on a function indicates its derivative with respect to its argument and the final identity is obtained on recognizing that $x_{mn}$ is a zero of $J_m(\cdot)$ and at such a zero, $J_m^\prime (x_{mn})=-J_{m+1}(x_{mn})$.

Having evaluated the inner product (\ref{FBKr}) in closed form in the above manner, we may now calculate the probabilities of detecting the imaging photon emitted in the state given by the SPDO (\ref{SPDO1}) in the various FB modes in a manner analogous to that used for the Zernike modes. The partial derivatives of these probabilities with respect to the point-source polar coordinates $r_0,\phi_0$ are also evaluated analogously. From these probabilities and their partial derivatives, we may calculate the FI for estimating the radial and azimuthal coordinates of the point source and the hole center for our two problems using expression (\ref{FIMatEl}). 

\subsection{Localized-Source Modes}
\begin{figure}
\centerline{
\includegraphics[width=0.4\textwidth]{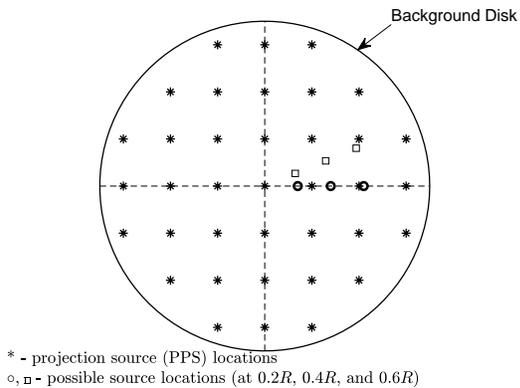}}
\caption{A diagram showing the locations of PPSs (*'s) and six possible locations of the point source (o's and \scalebox{0.6}{$
\square$}'s) being localized inside the background disk}
\label{fig:LocModes}
\end{figure}

We next considered projection modes that are constructed out of localized point-source wavefunctions by a Gram-Schmidt orthogonalization (GSO) procedure \cite{Gander80,Bjork94}. These sources, which we call projection point sources (PPSs), were at first chosen to be located on a square grid inside the uniformly illuminated background disk at positions $(-R+(m-1/2)a,-R+(n-1/2)a)$, with $m,n=1,\ldots, N_a$ and $a$, the grid spacing along each Cartesian dimension, being equal to $a=2R/N_a$, and keeping only those PPSs that are located inside the disk. This is shown in Fig.~\ref{fig:LocModes} for $N_a=7$.
Let us label the PPS locations serially by the vector $\br_i,\ i=1,\ldots, N_s$, in which $N_s\sim N_a^2$ is the total number of PPSs, and the corresponding unit-norm state vectors as $\ket{K_i},\ i=1,\ldots,N_s$. The latter are in general non-orthogonal and thus must be orthogonalized first by the GSO procedure. This is most efficiently performed by a Cholesky factorization of the corresponding $N_s\times N_s$ Gram matrix, $\bG$, of elements $G_{ij}\defeq\braket{K_i}{K_j}$, into a product of a lower triangular matrix and its transpose, 
\be
\label{cholesky}
\bG=\bL\bL^T. 
\ee
The orthonormalized set of projection modes, $\{\ket{e_j}|j=1,\ldots,N_s\}$, are linear combinations of the PPS states that may be expressed by the matrix relation,
\be
\label{GSmatrixrel}
\begin{pmatrix}
\bra{e_1}\\ \bra{e_2} \\ \vdots \\ \bra{e_{N_s}}
\end{pmatrix}
=\bC
\begin{pmatrix}
\bra{K_1}\\ \bra{K_2} \\ \vdots \\ \bra{K_{N_s}}
\end{pmatrix},
\ee
in which the coefficient matrix $\bC$ is a lower triangular matrix. Multiplying Eq.~(\ref{GSmatrixrel}) to its right by its Hermitian adjoint row vector and noting that the resulting matrix on the LHS is the identity matrix because of the orthonormality of the sought states $\ket{e_n}$, while the column-row vector product on the RHS yields the Gram matrix, with Cholesky factorization (\ref{cholesky}), we obtain the relation,
\ba
\label{Ortho}
\bC\bG\bC^T&={\bf{E}}\nn
{\rm i.e.,}\ \bC\bL(\bC\bL)^T&={\bf E},
\end{align}
in which ${\bf E}$ is the $N_s\times N_s$ identity matrix. In other words, the product, $\bC\bL$, yet another lower triangular matrix, must be a real orthogonal matrix with orthonormal rows, a fact that allows for an efficient row-by-row solution for $\bC$ via Gauss elimination. 

The classical GS orthogonalization via Cholesky factorization that we have presented here becomes numerically unstable when we include more than 35-40 PPS sources. More robust, modified GS procedures that more efficiently mitigate round-off errors may then be needed \cite{Bjork94}. 

The mode projection probabilities may be expressed as
\ba
\label{ModeProb}
P_n=&\mel{e_n}{\varrho}{e_n}\\
      =&\sum_{i,j=1}^n C_{ni} \mel{K_i}{\varrho}{K_j}C_{nj},
\end{align}
which can be evaluated numerically for the two SPDOs  (\ref{SPDO1}) and (\ref{H_SPDO}) we have considered in the paper. By differentiating these probabilities with respect to $r_0$ and $\phi_0$ and using expression (\ref{FIMatEl}), we may then calculate the FI for localizing the point source or hole for the two problems in the PPS basis. Note, however, that since the number of PPS modes is finite, the PPS basis is incomplete. For a correct evaluation of the FI matrix, we must therefore add to the sum (\ref{FIMatEl}) an extra term of the same form as the summand, corresponding to all the modes in which a photon is {\em unobserved}, which has the probability $\bar P=1-\sum_{n=1}^{N_s} P_n$ with partial derivatives $\pmu \bar P=- \sum_{n=1}^{N_s}\pmu P_n$. 

\subsection{Cram\'er-Rao bounds for Direct Imaging and Zernike, FB, and Localized-Mode Projections}

The diagonal elements of the inverse of the $2\times2$ FI matrix represent the two CRBs for unbiased estimation of the radial and arc-length coordinates of either the point source or the center of the tiny hole in the background disk.  We next discuss the numerically evaluated CRBs for each problem in the three wavefront projection bases that we have considered here. For the point-source localization problem, we plot the CRBs for direct imaging as well and compare them with the corresponding results for the three projection bases, but not for the hole problem as the comparisons for it are quite similar.

\subsubsection{\bf A point source in a uniformly illuminated disk}

We first display the CRB, as a function of the background-disk brightness parameter $b$, for estimating the radial coordinate of the point source using direct imaging with a clear aperture and each of the three projection basis sets we discussed in the previous three subsections. The PPS sources were chosen to be located in a regular square array with the central one of these sources being at the disk center and separated successively along the two Cartesian axes by a distance $a=R/3.5$. That allowed for a total of 37 PPS sources to be located inside the background disk, with all other sources that lie outside having been excluded from our projections. These are comprised of three central rows of 7 sources each, two rows of 5 sources each, and two outer rows of 3 sources, as shown in Fig.~10. The six possible locations of the point source being localized, three along the $x$ axis and three along the radial direction at angle $\pi/8$ with respect to the $x$ axis at distances $0.2R$, $0.4R$, or $0.6R$ from the disk center, are shown by $\circ$'s and \scalebox{0.6}{$\square$}'s on the same plot. In Figs.~11-14, we only show results for $\phi_0=\pi/8$, noting that the results for $\phi=0$ or any other values of $\phi_0$ are unremarkably similar.

Figure 11 displays the CRBs for a sub-diffractive disk of radius $R=0.5$, with two different values of the source radial distance $r_0$, either $0.2R$ (shown by x's) or $0.6R$ (shown by {\small $\Diamond$}'s). The CRBs for DI and all three wavefront projection bases approach the quantum CRB (QCRB) given by the corresponding diagonal element of the inverse of the QFI matrix, in the limit of vanishing background, $b\to 0$. That is not true for the larger-radius background disk, $R=2$, however, for which the same CRBs are shown in Fig.~12. Only for $r_0/R=0.2$ or smaller, for which the source is 0.4 units or less away from the center, does there seem to be a convergence of the Zernike-based CRB and QCRB values in this limit. This is consistent with our previous work \cite{YuPrasad18} on pair superresolution in which we showed that Zernikes fail to constitute an optimal basis when the pair separations are not much smaller than the Raylegh diffraction scale, of order 1 in the units we have used. Arguably, the FB and PPS modes are no different in this respect. 

\begin{figure}
\centerline{
\includegraphics[width=0.55\textwidth]{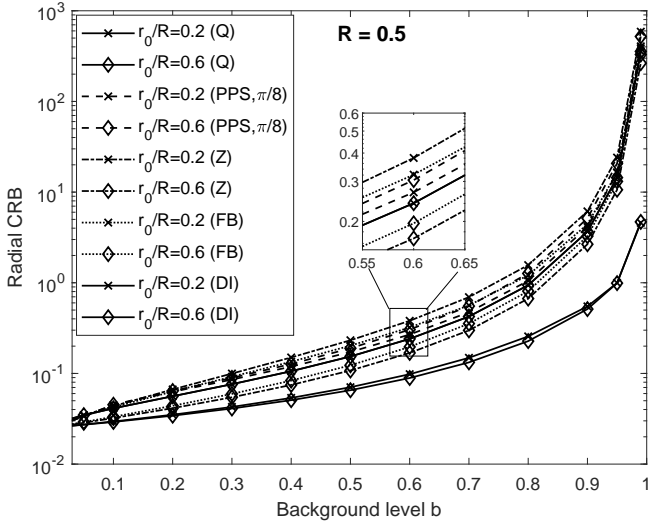}}
\caption{Plots of CRB for estimating $r_0$  vs. $b$ for direct imaging (solid upper curves) and for each of our three projection bases, for disk radius $R=0.5$ and for two different values of $r_0/R$. The quantum CRBs are shown by solid lines (lower solid curves).}
\end{figure}

\begin{figure}
\centerline{
\includegraphics[width=0.55\textwidth]{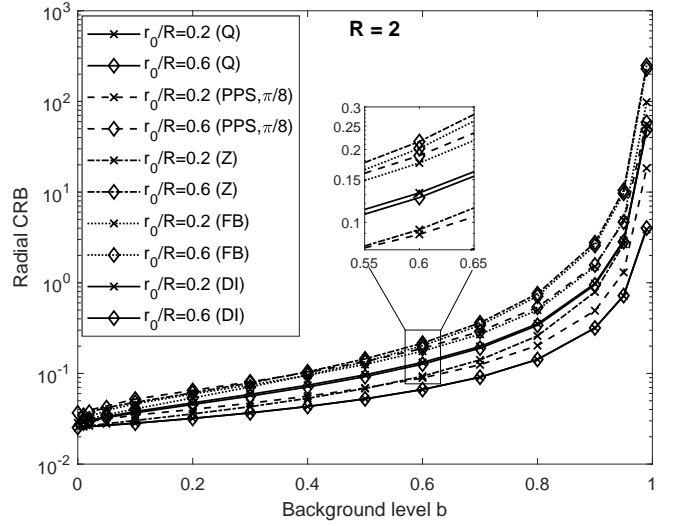}}
\caption{Same as Fig.~11 except that $R=2$.}
\end{figure}

Very similar behaviors are seen for the CRBs for estimating the orthogonal, azimuthal arc-length coordinate of the source position, namely $r_0\phi_0$, as well, as shown in Figs. 13 and 14. But estimating the arc-length coordinate of the source entails higher CRB values than estimating its radial coordinate, particularly at highly sub-diffractive scales and $b$ significantly different from 0, as seen in the highest of the dashed curves in Fig.~13 for which the point source is only a tenth of a unit away from the background disk center.

\begin{figure}
\centerline{
\includegraphics[width=0.55\textwidth]{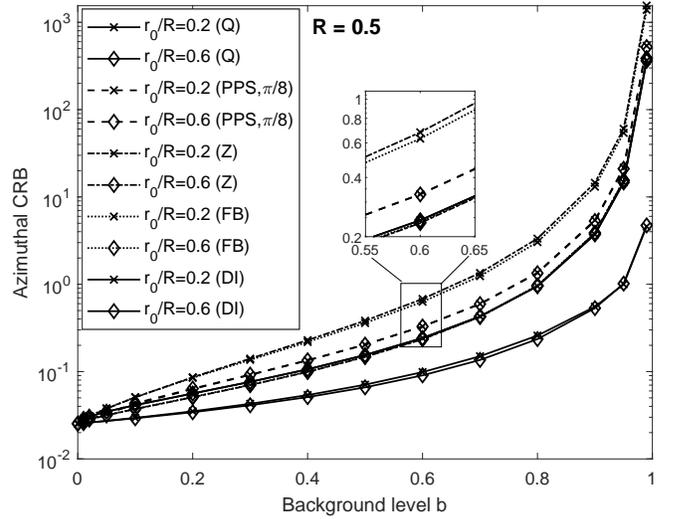}}
\caption{Plots of CRB for estimating the azimuthal arc-length coordinate, $r_0\phi_0$ vs. $b$ for direct imaging (upper solid curves) and each of our three projection bases, for disk radius $R=0.5$ and for two different values of $r_0/R$. The quantum CRBs are shown by solid lines (lower solid curves).}
\end{figure}

\begin{figure}
\centerline{
\includegraphics[width=0.55\textwidth]{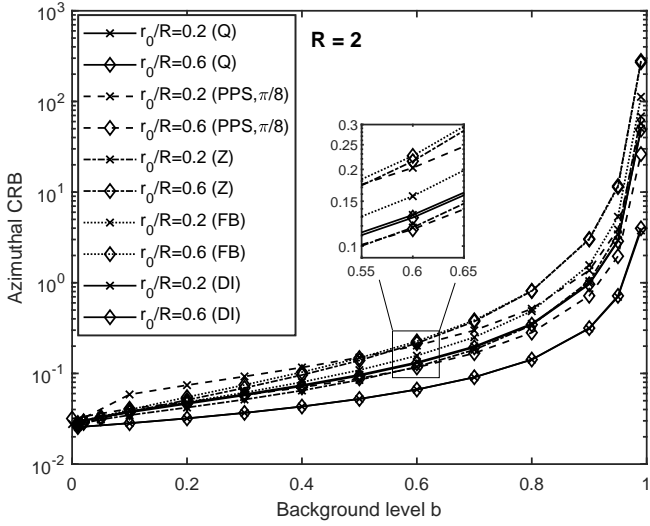}}
\caption{Same as Fig.~13 except that $R=2$.}
\end{figure}

Note that both the radial and azimuthal QCRBs take values that are essentially indistinguishable, at each value of $b$, regardless of the radial distance of the point source, particularly for the larger-radius disk for which $R=2$. Also, all CRBs, including the QCRBs, rise in value with rising background level, which is expected since a rising background strength makes the source localization more noisy, but the gap between the projection-CRB curves for each basis and the QCRB curves also becomes increasingly larger. All three projection bases become less and less efficient in estimating the point-source coordinates with rising background brightness levels. The zoomed-in insets on Figs.~11-14 magnify the differences between the various CRBs in the mid-range of the values for $b$ for an easier visual comparison between DI CRBs and the CRBs for the three wavefront projection bases.
\begin{figure}[htb]
\centering
\subfloat[]
{\includegraphics[width=0.55\textwidth]{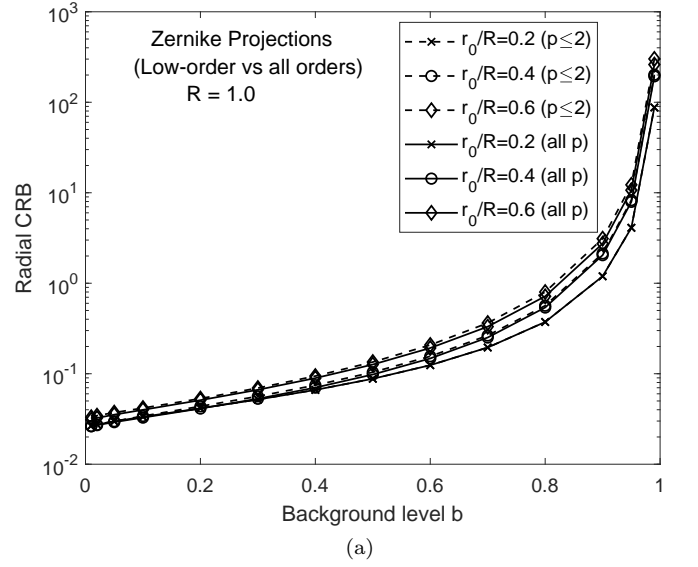}}\\
\subfloat[]
{\includegraphics[width=0.55\textwidth]{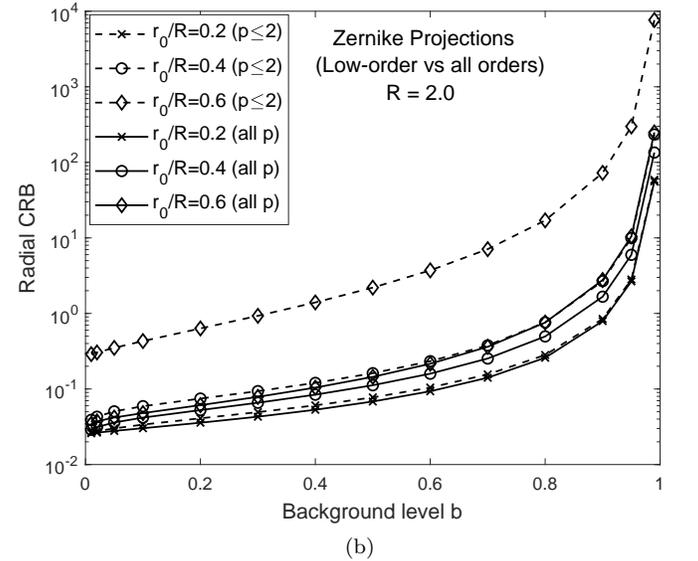}}
\\
\caption{\label{fig:ZernikeRadLoc_Finite_vs_InfiniteOrder} Plots of the CRB for radial localization of the point source vs. the brightness level of the background disk for (a) $R=1$ and (b) $R=2$, when only modes to quadratic order and when essentially all modes are included. }
\end{figure}

All three mode projections perform rather similarly when estimating the two coordinates of the point source against a background disk that has sub-diffractive extensions, {\it i.e.,} for $R<1$, although the PPS modes perform rather uniformly with respect to changing radial distance of the source, $r_0$, as seen in the pair of dashed lines tracking each other closely over the entire range of background brightness level. But for the larger disk, for which $R=2$, two differences may be noted. First, it is the FB modes that have a more uniform performance with changing radial distance of the source within that disk. Second, the PPS modes seem to perform the best, as seen in the dashed lines being lower, in each case of $r_0/R=0.2$ and $r_0/R=0.6$, than the dot-dashed and dotted lines corresponding to the Zernike and FB modes, particularly when the background is significantly bright with $b\geq 0.4$. But the gaps between the various CRBs and QCRBs remain finite and increase with increasing fractional background level $b$. None of the bases is thus an optimal basis at the quantum single-photon level when the background constitutes a significant fraction of the overall luminosity.

We note here that any practical implementation of the Zernike and FB mode projection protocols would employ only a finite number of modes of the lowest few orders with the rest of the higher-order modes that complete these bases remaining unobserved. The use of even fewer than ten lower-order modes of wavefront projection, as we have seen in our numerical computations, tends to achieve a pretty significant fraction of the total information about the location coordinates attained by each complete basis, at least in the sub-diffractive regime. We show this result in Figs.~15 (a) and (b) where we plot for $R=1$ and $R=2$, respectively, the radial-localization CRBs vs. the background brightness level, comparing the results for projections into the 8 lowest-order Zernike modes, corresponding to $p=0,1,2$ in Eq.~(\ref{Zprob}), with those obtained when essentially all Zernike modes (out to orders as high as $p=60$) are included in the projections. For $R=1$ there is little gain when including projections beyond those for which $p\leq 2$, while for $R=2$, the differences become quite pronounced only for the larger radial source distances, as seen here for $r_0/R=0.4$ and 0.6. Similar comparisons were obtained for azimuthal localization and for the FB and PPS projection bases as well, and were also discussed in our previous work \cite{Prasad20b,Prasad20c} for other parameter-estimation problems. %We will return to a study of the quantitative changes in the CRBs with increasing mode number for our final set of projection modes, by its very construction a finite set, in the context of the brightness-hole-localization problem. 

Figures 11-14 also show that direct imaging CRBs are on the whole not distinctly worse than the CRBs for the three wavefront projection bases discussed here. One must remember, however, that our DI CRB calculations have assumed observations that are neither pixelated nor suffer from any readout noise, which is unrealistic. Pixelation, in particular, will be increasingly more important and limiting of localization precision in practice \cite{Rieger14}, the more sub-diffractive the background disk happens to be. By contrast, the wavefront projection (WFP) technique could be designed to record the projection data on single camera pixels, one pixel per mode, for which only a few pixels would suffice, since, as we have noted in Figs.~15 (a) and (b), only a few modes of each basis are needed for closely approaching the corresponding complete-basis CRB values. The practical overhead of pixelation induced resolution loss and detection noise is thus expected to be greatly mitigated for the WFP technique. 

 \subsubsection{\bf A tiny hole in an otherwise uniformly illuminated disk}
 
We next consider our second problem for which we computed the CRB for estimating the location of the center of a tiny hole in an otherwise uniformly illuminated disk using wavefront projections in the same three bases.  As for the source localization problem, we use wavefront projections into {\em all} of the basis functions of the Zernike and FB mode bases, but only a handful of the PPS modes. For the last basis, we used two different sets of modes, one with 21 randomly placed PPS sources inside the background disk and the other containing 10 additional randomly placed sources for a total of 31 PPS sources, to show how adding more sources improves the estimation error for localizing the hole. The locations of these PPS sources are shown in Fig.~16, with the first 21 marked by x's and the additional 10 by *'s. Note that unlike the use of uniformly distributed PPS sources for the problem of source localization against the background disk that we discussed earlier, we have chosen the PPS locations randomly here. We will see from our results that the choice of PPS locations is largely immaterial. In fact, the GS orthogonalization of the PPS wavefunctions delocalizes their footprint, largely negating any {\em a priori} biases that might inadvertently be introduced into the estimation by a specific choice of the PPS locations, as long as there are sufficiently many of them distributed more or less uniformly over the full disk. We will not discuss here any comparisons of direct-imaging CRB values with the wavefront-projection CRB values for the brightness-hole problem, as they are quite analogous to the comparisons we discussed earlier for the source-localization problem. 
\begin{figure}
\centerline{
\includegraphics[width=0.45\textwidth]{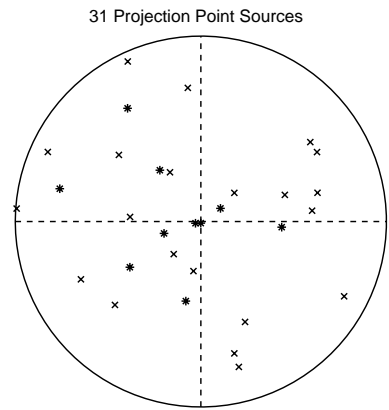}}
\caption{A diagram showing the locations of all 31 PPS sources with 21 locations shown by x's and 10 by *'s.}
\label{fig:LocModes2}
\end{figure} 

In Fig.~17, we display the CRB, scaled by the same factor $\epsilon$ by which the corresponding QFI values were inversely scaled in Figs.~6 and 7, for estimating the radial distance, $r_0$, of the hole center from the disk center as  a function of $r_0/R$ for two different values of the disk radius, 0.5 and 1.5. Here the PPS basis contained 21 sources. Figure 18, on the other hand, presents the similarly scaled CRB for estimating the azimuthal arc length of the hole position for the same three projection bases. As we can see, the Zernike (dashed curves) and FB (dot-dashed curves) mode projections perform rather similarly and somewhat worse than PPS mode projections, particularly for the larger disk radius ({\tiny $\square$} markers). The superiority of the PPS basis is even more pronounced for both disk radii considered here over much of the $r_0/R$ range for azimuthal arc length estimation, as we can see from Fig.~17. This advantage can be even greater in a practical setting where only finite numbers of Zernike or FB modes can be included on any projection protocol.  
\begin{figure}
\centerline{
\includegraphics[width=0.55\textwidth]{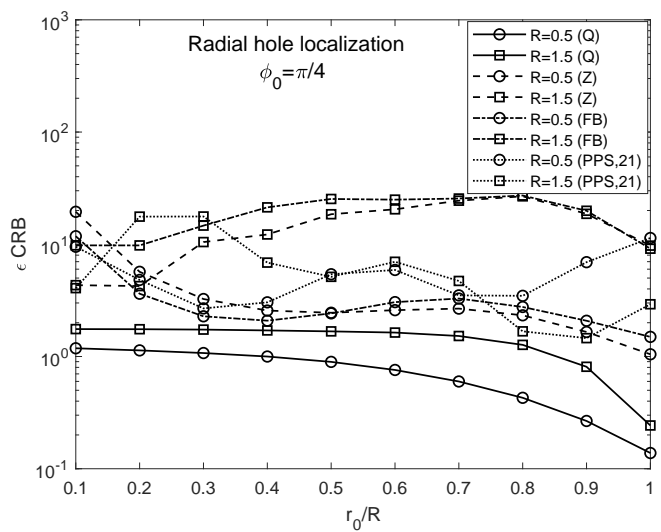}}
\caption{Plots of CRB for estimating $r_0$ for the Zernike (dashed curves), FB (dott-dashed curves), and 21-mode PPS (dotted curves) basis sets. The CRB has been scaled by factor $\epsilon=(\delta_0/R)^2$, the ratio of hole area to the disk area. The similarly scaled quantum CRBs are shown by solid lines.}
\end{figure}

\begin{figure}
\centerline{
\includegraphics[width=0.55\textwidth]{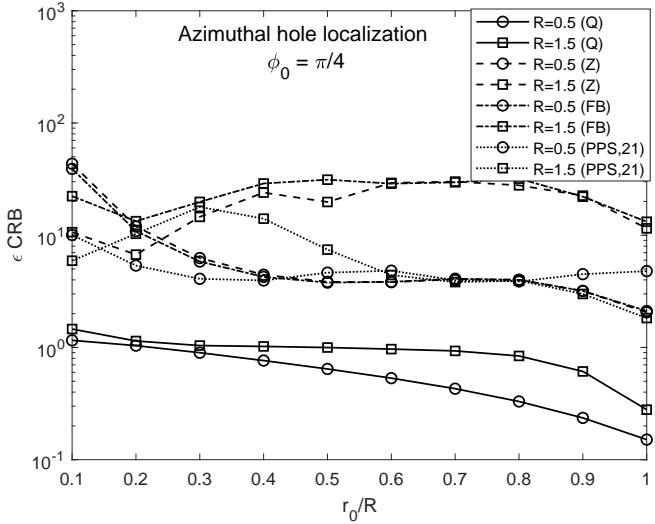}}
\caption{Same as Fig.~18 except for azimuthal localization}
\end{figure}

In Figs.~19 and 20, we demonstrate the improvement of the CRB for estimating the radial and azimuthal arc length coordinates of the brightness hole center when we include 10 additional PPS sources into the PPS projection set. We can see that for all three disk radii considered here, the reduction of the CRB with additional PPS sources is rather minimal across the full range of possible radial locations of the hole. The largest improvement seems to result for either the largest disk radius ({\tiny $\square$} markers) or for hole locations closer to the perimeter for the smallest, subdiffractive disk radius of 0.5 ({$\circ$} markers). Although we do not show the CRB values for the case of modes formed from 37 uniformly located PPS sources in the geometry shown in Fig.~10, they are quite comparable for both radial and azimuthal CRBs. 
\begin{figure}
\centerline{
\includegraphics[width=0.55\textwidth]{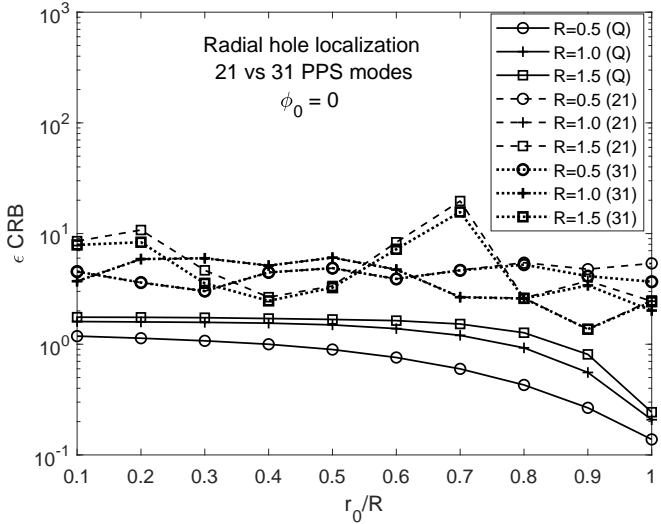}}
\caption{Plots of CRB for estimating $r_0$ of the hole center for the localized PPS modes, $\phi_0=0$, and three values of $R$, with the dashed lines showing results for 21 PPS sources and dotted lines for 31 PPS sources, as indicated in parentheses in the legend.}
\end{figure}

\begin{figure}
\centerline{
\includegraphics[width=0.55\textwidth]{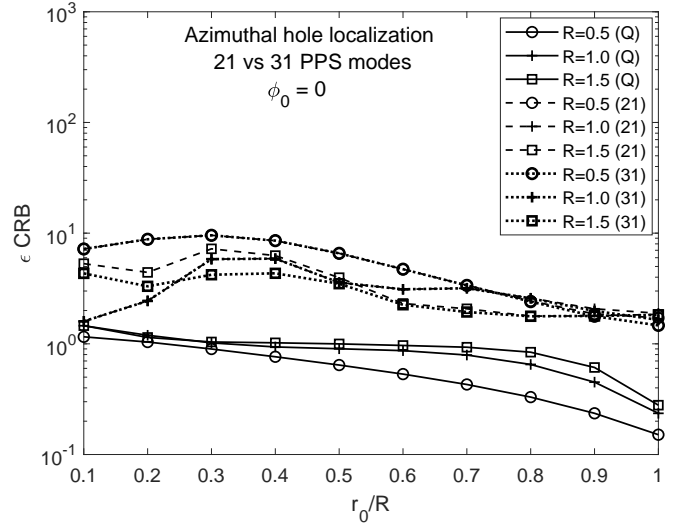}}
\caption{Same as Fig.~19 except for azimuthal localization}
\end{figure}

%As for the case of locating a point source in a uniform background disk, the CRB for locating a hole in such a disk is rather similar for the Zernike and FB sets of projection modes, neither set being particularly optimal with respect to the corresponding QCRB. The orthogonalized PPS modes perform distinctly better, however, as we see from Figs.~26 and 29, with the corresponding CRBs one to two orders smaller and typically within one order of the quantum bound on the CRB over the full range of the distance of the hole from the disk center. Further, even with 21 orthogonalized modes formed out of 21 randomly positioned PPS sources, the CRBs, shown by dashed lines,  are already quite small, with only minor further reductions resulting from the inclusion of 10 additional PPS modes, as shown by dotted lines. 
The finite gaps between the wavefront-projection CRBs and the corresponding QCRBs seem to remain, however, regardless of the specifics of the three projection basis. It is quite possible that QCRBs are either fundamentally unattainable or, in more practical terms, unattainable with wavefront projections alone. One may need weaker quantum measurements, those that are only describable in terms of general Kraus operators and a non-projective POVM, to be able to fully bridge, or at least further reduce, the gap between the measurement CRB and QCRB. But just what sort of quantum processess and estimators one might need to achieve this remains an open question.   

\section{Concluding Remarks}

We calculated the lower bounds on the variance for estimating the 2D location coordinates of two different kinds of point object in a uniformly bright circular disk of fixed center and radius, that point object being either a point source or a brightness hole located somewhere inside the disk. Specifically, we computed the QFI matrix and its inverse matrix, the diagonal elements of which yield the lowest possible variances of any unbiased, simultaneous estimation of these coordinates, namely the QCRBs. In the limit that the point source being localized is relatively bright compared to the background disk, a perturbative analysis of the SLDs suffices, as we showed by means of a detailed numerical evaluation of the exact results, to generate accurate expressions for QFI and QCRB, In fact, the perturbative treatment was found to maintain excellent accuracy even when the integrated brightness of the background disk was comparable to the brightness of the point source.   

For the hole-in-the-disk problem, the perturbative approach was all that was needed when the hole is small in size compared to the disk radius, as we assumed, and can thus be treated as a point brightness deficit in the disk, with the ratio of the hole and disk areas serving as the small parameter. Our perturbative approach, nevertheless, required a numerical evaluation of the eigenvalues and eigenfunctions of the single-photon density operator of emission from a uniformly bright disk. Note also that a slight parametric modification of the contrast parameter, $\epsilon\to \epsilon(1-f)$, allows one to treat the problem of localizing a smaller unresolved star against a larger background star, as, e.g., in a binary star system, when the former is a fraction $f$ as bright per unit area as the latter.   

We next considered the attainability of the QCRB by means of wavefront projections for a variety of operating conditions for both problems, which included varying the fractional background disk brightness level, the 2D location of the source from the disk center, and the disk radius.  Three different sets of modes, specifically the Zernike, Fourier-Bessel and projection-point-source modes, the last with varying numbers of projection point sources, were employed for this purpose, and their relative efficiency in reaching the ultimate quantum bound was numerically assessed in detail. 

None of these bases was found to be particularly optimal for the source-localization problem, with all three bases yielding CRBs that only approach the QCRB within a factor of 1-10 over a wide range of values of the disk radius, source location, and background levels. A detailed comparison of the WFP based CRBs with DI based CRBs showed that wavefront projections are not significantly superior to direct imaging for source localization, which in the context of DI is largely a PSF fitting, rather than a superresolution, problem for which DI can achieve precisions of a few nm even at optical wavelengths \cite{Yildiz03}. We expect WFP protocols to be qualitatively superior to DI, however, when estimating in the photon-counting limit the pair separation vectors of two or more closely spaced sources whose brightness centroid is either known or can be well localized \cite{Tsang16,PrasadYu19}. 

For the hole-in-the-disk problem, the PPS modes performed the best among our three WFP bases, being a factor 2-10 better than the other projection sets in terms of their respective CRBs. It is quite possible that the QCRB is unattainable \cite{D-D20} in any {\em strong} projective measurement basis, and a specific {\em weak} measurement protocol involving an ancilla might be needed to yield a CRB that reaches the corresponding QCRB under controlled operating conditions. Or perhaps, {\em no} measurement POVM exists that is efficient in this sense, that being even more likely to be true when imperfect, noisy detection \cite{Len22} further reduces estimation efficiency. 

\acknowledgments
The author has enjoyed helpful communication with Jonathan Owens, James Bray, and Bogdan Neculaes of GE Research regarding potential applications of this work. 

\appendix
\section{Perturbative Expressions for QFI to Quadratic Order}

We first note the identity,
\be
\label{K0pnuK0}
\mel{K_0}{\pnu}{K_0}=\pnu\!\braket{K_0}{K_0}=0,
\ee
that follows readily from the inversion symmetry of the circular aperture.  Further, by expanding the exponential operator in the Lyapunov solution (\ref{Lyapunov}) in a power series and noting that $\varrho_0^n=\varrho_0=\ketbra{K_0}{K_0}$ for all $n\ge 1$, we may reduce it to the simpler form, 
\ba
\label{ExpFullRankDO}
\exp(-\lambda\varrho_\eta)=&\exp(-\lambda\eta/D)\nn
&\times \left\{\cI_N+\exp[-\lambda(1-\eta)]\varrho_0\right\},
\end{align}
in which 
\be
\label{IdN}
\cI_N=\cI-\varrho_0
\ee
defines the identity operator in the null space of the rank-1 point-source SPDO, $\varrho_0$. 

To evaluate the zeroth-order term, $K^{(0)}_{\mu\nu}$, in the expression for the QFI matrix elements given by Eq.~(\ref{Kn}), we find it simplest to evaluate $\cL^{(0)}_\nu$ first using the Lyapunov solution (\ref{Lyapunov}).  Substituting the expansion (\ref{ExpFullRankDO}) for the exponential operators on either side of $\pnu\varrho_0$ in Eq.~(\ref{Lyapunov}) and distributing the product of operators into four terms, we easily see that two of them must vanish since
\be
\label{Identities}
\cI_N\pnu\,(\!\ketbra{K_0}{K_0})\cI_N=\ketbra{K_0}{K_0}\pnu\,(\!\ketbra{K_0}{K_0})\!\ketbra{K_0}{K_0}=0,
\ee
identities that follow from a substitution of definition (\ref{IdN}) for $\cI_N$, the product identity, $\pnu\,(\!\ketbra{K_0}{K_0})=\pnu\!\ketbra{K_0}{K_0}+\ket{K_0}\pnu\!\bra{K_0}$, and the symmetry conditions (\ref{K0pnuK0}).  The remaining two terms involve simple exponential integrals that can be done in the limit $\eta\to 0^+$, leading to the result,
\ba
\label{cL0nu}
\cL^{(0)}_\nu = &2(\ket{K_0}\pnu\!\bra{K_0}\cI_N+\cI_N\pnu\!\ket{K_0}\bra{K_0})\nn
                     = &2(\ket{K_0}\pnu\!\bra{K_0}+\pnu\!\ket{K_0}\bra{K_0}),
\end{align}
with the second equality following from the fact that $\ket{K_0}\pnu\!\bra{K_0}\cI_N=\ket{K_0}\pnu\!\bra{K_0}$ according to definition (\ref{IdN}) for $\cI_N$ and identity (\ref{K0pnuK0}). Taking the matrix element of expression (\ref{cL0nu}) between  $\pmu\bra{K_0}$ and $\ket{K_0}$ and using identity (\ref{K0pnuK0}) once again generates, according to Eq.~(\ref{Kn}), the zeroth-order QFI matrix elements,
\be
\label{K0}
K_{\mu\nu}^{(0)}=4\Re\pmu\!\bra{K_0}\pnu\!\ket{K_0}.
\ee

Analogous to the Lyapunov solution (\ref{Lyapunov}) for the first of the relations in Eq.~(\ref{SLDp}), the second relation there, for $n\ge 1$, may be cast in the Lyapunov form as
\ba
\label{SLDn}
\cL^{(n)}_\nu=&-\lim_{\eta\to 0^+}\int_0^\infty \!\!d\lambda\, \exp(-\lambda\varrho_\eta)\nn
&\times\left(\cL_\nu^{(n-1)}\varrho_B+\varrho_B\cL^{(n-1)}_\nu\right)\exp(-\lambda\varrho_\eta).
\end{align}
Let us now substitute expansion (\ref{ExpFullRankDO}) for the pre and post exponential operators in Eq.~(\ref{SLDn}) and distribute the product inside the $\lambda$ integrand into four terms of form,
\be
\label{4terms}
\cI_N\cO\cI_N, \ \cI_N\cO\varrho_0,\ \varrho_0\cO\cI_N, \ \varrho_0\cO\varrho_0,
\ee
where $\cO=\cL_\nu^{(n-1)}\varrho_B+\varrho_B\cL^{(n-1)}_\nu$. The first of these terms must vanish identically, since from the second of Eqs.~(\ref{SLDp}) $\cO$ is also equal to $-(\cL_\nu^{(n)}\varrho_0+\varrho_0\cL^{(n)})$ and $\varrho_0\cI_N=\cI_N\varrho_0=0$. The remaining three terms in Eq.~(\ref{4terms}) yield simple exponential integrals over $\lambda$ that converge in the limit $\eta\to 0^+$ and are easily calculated, and Eq.~(\ref{SLDn}) thus reduces to the recursion relation,
\ba
\label{SLDn1}
\cL^{(n)}_\nu=&-\left[\cI_N(\varrho_B\cL^{(n-1)}_\nu +\cL^{(n-1)}_\nu\varrho_B)\varrho_0+\ha\right]\nn
                       &-{1\over 2}\varrho_0(\varrho_B\cL^{(n-1)}_\nu +\cL^{(n-1)}_\nu\varrho_B)\varrho_0\nn
                     =&-\left[(\varrho_B\cL^{(n-1)}_\nu +\cL^{(n-1)}_\nu\varrho_B)\varrho_0+\ha\right]\nn
                       &+{3\over 2}\varrho_0(\varrho_B\cL^{(n-1)}_\nu +\cL^{(n-1)}_\nu\varrho_B)\varrho_0,
\end{align}
in which definition (\ref{IdN}) was substituted in the first equality to reach the second equality.

Taking the matrix element of expression (\ref{SLDn1}) between $\pmu\bra{K_0}$ and $\ket{K_0}$ and using definition (\ref{IdN}) and identity (\ref{K0pnuK0}) to see that 
\be
\label{cIN1}
\pnu\bra{K_0}\cI_N =\pnu\bra{K_0}, \ \pnu\bra{K_0}\varrho_0=0,
\ee
yields, via relation (\ref{Kn}), the $n$th order correction to the QFI matrix elements as
\ba
\label{Kmunu_n}
K_{\mu\nu}^{(n)}=-2\Re\, &\pmu\!\bra{K_0}\big[\cL_\nu^{(n-1)}\varrho_B
+\varrho_B\cL^{(n-1)}_\nu\big]\ket{K_0},\nn
&\qquad\qquad n=1,2,\ldots
\end{align}

The first-order correction to QFI matrix elements follows immediately from relation (\ref{Kmunu_n}), for $n=1$, solution (\ref{cL0nu}), and identity (\ref{K0pnuK0}),
\ba
\label{K1}
K^{(1)}_{\mu\nu} =&-4\Re\big[\pmu\!\bra{K_0}\pnu\!\ket{K_0}\mel{K_0}{\varrho_B}{K_0}\nn
                                    &\qquad\qquad+\pmu\!\bra{K_0}\varrho_B\pnu\!\ket{K_0}\big].
\end{align}

To evaluate the second-order correction, $K^{(2)}_{\mu\nu}$, we need, as we see from relation (\ref{Kmunu_n}), $\cL^{(1)}_\nu\ket{K_0}$ and the Hermitian adjoint of $\cL^{(1)}_\nu\pmu\!\ket{K_0}$. Using solution (\ref{SLDn1}) for $n=1$, we may express the first of them as
\ba
\label{cL1nuK0}
\cL^{(1)}_\nu\ket{K_0}&=-(\varrho_B\cL^{(0)}_\nu +\cL^{(0)}_\nu\varrho_B)\ket{K_0}\nn
                                    &+{1\over 2}\varrho_0(\varrho_B\cL^{(0)}_\nu +\cL^{(0)}_\nu\varrho_B)\ket{K_0}\nn 
					 &=-2\varrho_B\pnu\!\ket{K_0}-\cL^{(0)}_\nu\varrho_B\ket{K_0}\nn
					  &+\ketbra{K_0}{K_0}\varrho_B\pnu\!\ket{K_0}+\ket{K_0}\pnu\!\mel{K_0}{\varrho_B\pnu}{K_0}\nn
					 &= -2\varrho_B\pnu\!\ket{K_0}-\cL^{(0)}_\nu\varrho_B\ket{K_0}\nn
					  &+2\ket{K_0}\Re(\bra{K_0}\varrho_B\pnu\!\ket{K_0}),
\end{align}
in which we used the identities $\varrho_0\cL^{(0)}_\nu=2\ket{K_0}\pnu\!\bra{K_0}$ and $\cL^{(0)}_\nu\ket{K_0}=2\pnu\!\ket{K_0}$ that follow from the solution (\ref{cL0nu}) in the first equality to arrive at the second equality and the fact that $z+z^*=2\Re z$, where $z=\mel{K_0}{\varrho_B\pnu\!}{K_0}$, to reach the final equality. Applying solution (\ref{SLDn1}) for $n=1$ on $\pnu\!\ket{K_0}$, noting that $\varrho_0\pnu\!\ket{K_0}=0$ according to symmetry condition (\ref{K0pnuK0}), and substituting solution (\ref{cL0nu}) for $\cL^{(0)}_\nu$, we may derive the second quantity we need for $K_{\mu\nu}^{(2)}$, namely
\ba
\label{cL1nu_pmuK0}
\cL^{(1)}_\nu\pmu\!\ket{K_0}=&-\varrho_0(\varrho_B\cL^{(0)}_\nu +\cL^{(0)}_\nu\varrho_B)\pmu\!\ket{K_0}\nn
                                            =&-2\ket{K_0}\mel{K_0}{\varrho_B}{K_0}\pnu\!\bra{K_0}\pmu\!\ket{K_0}\nn
                                              &-2\ket{K_0}\pnu\!\bra{K_0}\varrho_B\pmu\!\ket{K_0}.
\end{align}

On using Eq.~(\ref{cL1nuK0}) and the Hermitian adjoint of Eq.~(\ref{cL1nu_pmuK0}) in Eq.~(\ref{Kmunu_n}), with $n$ set equal to 2, we may evaluate $K^{(2)}_{\mu\nu}$ fully as
\ba
\label{K2}
K^{(2)}_{\mu\nu}=&4\mel{K_0}{\varrho_B}{K_0}^2\Re (\pmu\!\bra{K_0}\pnu\!\ket{K_0})\nn
                           &+8\mel{K_0}{\varrho_B}{K_0}\Re (\pmu\!\bra{K_0}\varrho_B\pnu\!\ket{K_0})\nn
                           &+4\Re (\pmu\!\bra{K_0}\varrho_B^2\pnu\!\ket{K_0})\nn
                           &-4\Im (\pmu\!\mel{K_0}{\varrho_B}{K_0})\Im(\pnu\!\bra{K_0}\varrho_B\ket{K_0}),
\end{align}
where we combined two identical quantities to obtain the second term and then used the identity, $\Re(z_1z_2)-\Re(z_1)\Re(z_2)=-\Im(z_1)\Im(z_2)$, where $z_1=\pmu\!\mel{K_0}{\varrho_B}{K_0}$ and $ z_2=\pnu\!\mel{K_0}{\varrho_B}{K_0}$, to obtain the final term on the RHS.

\section{QFI Expressions for Localizing a Point Source in a Uniformly Bright Disk}
Taking the inner product of the eigenvalue equation (\ref{SPDO1eigenstate}) with the bra $\bra{K_\brp}$ and using relation (\ref{KrKrp}) generates the RHS of Eq.~(\ref{CoeffFnEqn}) and thus, since the eigenfunctions $C_\lambda(\br)$ can always be chosen to be real,
\be
\label{Krp_lambda}
\braket{K_\brp}{\lambda_i}=\braket{\lambda_i}{K_\brp}=\lambda_i C_i(\brp).
\ee
Multiplying the far left and far right sides of this equation, each on its right by $\bra{\lambda_i}$, adding over all $i$, and using the completeness of the full set of orthonormal eigenstates of the SPDO yields the sum rule,
\be
\label{sumrule1}
\bra{K_\brp}=\sum_{i\in\cS}\lambda_iC_i(\brp)\bra{\lambda_i},
\ee
in which the sum over all $i$ may be restricted to that only over the support, $\cS$, because of the factor $\lambda_i$ in the sum.

Multiplying Eq.~(\ref{CoeffFnEqn}) for the $i$th eigenfunction by $\lambda_i C_i(\br^{\prime\prime})$ on both sides and then summing over all $i$ implies the following relation:
\be
\label{I1I2}
\int G(\brp) {2J_1(2\pi|\br-\brp|)\over 2\pi|\br-\brp|}I_1(\brp,\br^{\prime\prime}) dA'=I_2(\br,\br^{\prime\prime}),
\ee
where $I_n(\br,\brp)$ is defined as the sum
\be
\label{In}
I_n(\br,\brp)\defeq\sum_i\lambda_i^nC_i(\br)C_i(\brp).
\ee
But from identity (\ref{Krp_lambda}) we see that $I_2(\br,\brp)$ is simply the sum of products $\braket{K_\br}{\lambda_i}\braket{\lambda_i}{K_\brp}$ over the complete set of orthonormal eigenstates $\{\lambda_i\}$. Thus, from the completeness relation for these eigenstates, it follows that 
\be
\label{I2}
I_2(\br,\brp)=\braket{K_\br}{K_\brp}.
\ee
Thus, in view of relation (\ref{KrKrp}), we may then express Eq.~(\ref{I1I2}) as
\be
\label{I1I2a}
\int G(\brp) {2J_1(2\pi|\br-\brp|)\over 2\pi|\br-\brp|}I_1(\brp,\br^{\prime\prime}) dA'={2J_1(2\pi|\br-\br^{\prime\prime}|)\over 2\pi|\br-\br^{\prime\prime}|}.
\ee
By inspection of the two sides of this equation, we immediately recognize that $G(\brp)\,I_1(\brp,\br^{\prime\prime})$ must be equal to the Dirac delta function, $\delta^{(2)}(\br-\br^{\prime\prime})$, {\em i.e.}, 
\be
\label{I1}
I_1(\br,\brp)={\delta^{(2)}(\br-\brp)\over G(\br)},
\ee
up to an additive function belonging to the null space of the kernel function, $2J_1(2\pi|\br-\brp|)/|\br-\brp|$. Our numerical calculations indicate no null space for the eigenstates of the disk SPDO, so we may assume Eq.~(\ref{I1}) to be exact. In the special case that $\br=\brp=\br_0$, using the defintion (\ref{Gdef}) for $G$, we may reduce this identity the value, 
\ba
\label{I10}
I_1(\br_0,\br_0)=&{\delta^{(2)}(0)\over (1-b)\delta^{(2)}(0)+b/(\pi R^2)}\nn
                        =&{1\over 1-b+b(\delta_0^2/R^2)}\buildrel {\delta_0\to 0} \over  \longrightarrow {1\over 1-b},
\end{align}
where we used the discretized form of the Dirac $\delta$ function singularity, $\delta^{(2)}(0)$,  corresponding to the interpretation that any point source, in practice, has a finite radial extension $\delta_0$ to replace $\delta^{(2)}(0)$ by $1/(\pi\,\delta_0^2)$.

Using results (\ref{Krp_lambda}), (\ref{I1}),
\be
\label{pmu_rho}
\pmu\varrho=(1-b)[\pmu\ketbra{K_0}{K_0}+\ket{K_0}\pmu\bra{K_0}],
\ee
$\braket{K_0}{K_0}=1$, and identity (\ref{K0pnuK0}), we may see that
\ba
\label{lam_pmurho_pnurho_lam}
\bra{\lambda_i}\pmu\varrho\ket{\lambda_j}=(1&-b)\big[\lambda_iC_i(\br_0)\,Q_{j\mu}+\lambda_jC_j(\br_0)Q_{i\mu}\big];\nn
\bra{\lambda_i}\pmu\varrho\, \pnu\varrho\ket{\lambda_i} =&(1-b)^2\big[Q_{i\mu}Q_{i\nu}\nn
&+\lambda_i^2C_i^2(\br_0)\,\pmu\bra{K_0}\pnu\ket{K_0}\big],
\end{align}
in which $Q_{i\mu}$ denotes the real matrix element that we have already evaluated in Eq.~(\ref{lambdai_pmu_K0}),
\be
\label{Qimu}
Q_{i\mu}\defeq \mel{\lambda_i}{\pmu}{K_0}=\pmu\braket{K_0}{\lambda_i}=Q_{\mu i}.
\ee
Dividing the second of the relations (\ref{lam_pmurho_pnurho_lam}) by $\lambda_i$ and summing over all $i$ yields the following expression for the first of the terms in expression (\ref{Hmunu1}):
\ba
\label{Hmunu_singlesum}
\sum_{i\in\cS}{1\over\lambda_i}&\bra{\lambda_i}{\pmu\varrho\,\pnu\varrho}\ket {\lambda_i}=(1-b)^2\Big[\pnu\bra{K_0}{\varrho^{(+)}\pmu}\ket{K_0}\nn
&+I_1(\br_0,\br_0)\pmu\bra{K_0}\pnu\ket{K_0}\Big],
\end{align}
in which $\varrho^{(+)}$ is the Moore-Penrose pseudo-inverse of the SPDO,
\be
\label{MPinv}
\varrho^{(+)}=\sum_{i\in \cS}{1\over\lambda_i}\ketbra{\lambda_i}{\lambda_i}.
\ee

Next, we use the first of the relations (\ref{lam_pmurho_pnurho_lam}) in the product of the matrix elements in the symmetric double sum in Eq.~(\ref{Hmunu1}), distribute out the resulting product into four terms, and subsequently interchage the $i$ and $j$ indices on two of the four terms, which makes them equal, pairwise, to the other two terms. We may thus express that symmetric double sum as
\ba
\label{Hmunu_doublesum}
&\sum_{i,j\in\cS}\left({1\over\lambda_i+\lambda_j}\!-\!{1\over\lambda_i}\!-\!{1\over\lambda_j}\right)\Re[\mel{\lambda_i}{\pmu\varrho}{\lambda_j}\!\!\mel{\lambda_j}{\pnu\varrho}{\lambda_i}]\nn
&=2(1-b)^2\sum_{i,j\in\cS}\left[{1\over\lambda_i+\lambda_j}\!-\!{1\over\lambda_i}\!-\!{1\over\lambda_j}\right]\nn &\times\big[\lambda_i\lambda_jC_i(\br_0)\,C_j(\br_0)Q_{i\mu}Q_{j\nu}\!+\!\lambda_i^2C_i^2(\br_0)Q_{j\nu}Q_{j\mu}\big]\nn
&=2(1-b)^2\nn
&\times\Bigg[\sum_{i,j\in\cS}{\lambda_i\lambda_jC_i(\br_0)\,C_j(\br_0)Q_{i\mu}Q_{j\nu}\!+\!\lambda_i^2C_i^2(\br_0)Q_{j\nu}Q_{j\mu}\over\lambda_i+\lambda_j}\nn
&\!-\!\sum_{i\in\cS}C_i(\br_0)Q_{i\mu}\sum_{j\in\cS}\lambda_jC_j(\br_0)Q_{j\nu}\!-\!I_1(\br_0,\br_0)\sum_{j\in\cS}Q_{j\nu}Q_{j\mu}\nn
&\!-\!\sum_{j\in\cS}C_j(\br_0)Q_{j\nu}\sum_{i\in\cS}\lambda_iC_i(\br_0)Q_{i\mu}\!-\!\sum_{j\in\cS}{Q_{j\nu}Q_{j\mu}\over\lambda_j}\Bigg],
\end{align}
where we used the definition (\ref{In}) for $n=1,2$ and the identity (\ref{I2}), for the special case of $\br=\brp=\br_0$ for which it is simply equal to 1, to arrive at the final expression. We next note three identities,
\ba
\label{Qidentities}
\sum_{j\in\cS}\lambda_jC_j(\br_0)Q_{j\nu}=&\sum_{j\in\cS}\lambda_jC_j(\br_0)\mel{\lambda_j}{\pnu}{K_0}\nn
                                                                =&\mel{K_0}{\pmu}{K_0}=0;\nn
\sum_{j\in\cS}Q_{j\nu}Q_{j\mu}=&\pnu\bra{K_0}\sum_{j\in \cS}\ketbra{\lambda_j}{\lambda_j}\pmu\ket{K_0}\nn
                                                 =&\pnu\bra{K_0}\cI_S\pmu\ket{K_0};\nn
\sum_{j\in\cS}{Q_{j\nu}Q_{j\mu}\over\lambda_j}=&\pnu\bra{K_0}\sum_{j\in \cS}{\ketbra{\lambda_j}{\lambda_j}\over\lambda_j}\pmu\ket{K_0}\nn
=&\pnu\bra{K_0}\varrho^{(+)}\pmu\ket{K_0},
\end{align}
where we used the sum rule (\ref{sumrule1}) to reach the second line of the first identity, which vanishes due to relation (\ref{K0pnuK0}), and the reality condition (\ref{Qimu}) on $Q_{j\nu}$ to write it as $\pnu\braket{K_0}{\lambda_j}$ to reach the first equality in the last two identities, the symbol
\be
\label{Is}
\cI_S=\sum_{i\in\cS}\ketbra{\lambda_i}{\lambda_i}
\ee
for the identity operator in the support subspace of $\varrho$, and definition (\ref{MPinv}) to reach the final equalities in these two latter identities. Substituting relations (\ref{Qidentities}) into Eq.~(\ref{Hmunu_doublesum}) greatly simplies it with the result,
\ba
\label{Hmunu_doublesum1}
&\sum_{i,j\in\cS}\left({1\over\lambda_i+\lambda_j}\!-\!{1\over\lambda_i}\!-\!{1\over\lambda_j}\right)\Re[\mel{\lambda_i}{\pmu\varrho}{\lambda_j}\!\!\mel{\lambda_j}{\pnu\varrho}{\lambda_i}]\nn
&=2(1-b)^2\nn
&\times\Bigg[\sum_{i,j\in\cS}{\lambda_i\lambda_jC_i(\br_0)\,C_j(\br_0)Q_{i\mu}Q_{j\nu}\!+\!\lambda_i^2C_i^2(\br_0)Q_{j\nu}Q_{j\mu}\over\lambda_i+\lambda_j}\nn
&\!-I_1(\br_0,\br_0)\pnu\bra{K_0}\cI_S\pmu\ket{K_0}-\pnu\bra{K_0}\varrho^{(+)}\pmu\ket{K_0}\Bigg],
\end{align}

When we substitute expressions (\ref{Hmunu_singlesum}) and (\ref{Hmunu_doublesum1}) into Eq.~(\ref{Hmunu1}) and perform some cancellations, we obtain the following final expression for the QFI matrix elements:
\ba
\label{Hmunu4}
H_{\mu\nu}&=4(1-b)^2 \Bigg[{1\over 1-b}\pnu\bra{K_0}\cI_N\pmu\ket{K_0}\nn
+&\sum_{i,j\in\cS}{\lambda_i\lambda_jC_i(\br_0)\,C_j(\br_0)Q_{i\mu}Q_{j\nu}\!+\!\lambda_i^2C_i^2(\br_0)Q_{j\nu}Q_{j\mu}\over\lambda_i+\lambda_j}\Bigg],
\end{align}
in which the symbol $\cI_N$ defines the identity operator in the null subspace of the SPDO,
\be
\label{cIN}
\cI_N=\sum_{i\in\cN}\ketbra{\lambda_i}{\lambda_i}=\cI-\cI_S,
\ee
and we used the value (\ref{I10}) for $I_1(\br_0,\br_0)$. As we have noted earlier, the point-source-background-disk SPDO that we are analyzing here appears to be full-rank, with the first term on the RHS of Eq.~(\ref{Hmunu4}) vanishing in our numerical calculations to an accuracy of roughly 1 part in 10$^5$. But for the sake of generality, we do not make this approximation in Eq.~(\ref{Hmunu4}), leaving it unchanged.

\section{QFI Expressions for Localizing a Hole in a Uniformly Bright Disk}  

Let us first recall certain identities from Ref.~\cite{Prasad20b} involving the eigenvalues and eigenstates of the uniformly-bright disk problem, which we need here:
\ba
\label{Disk_identities}
\braket{K_\br}{\lambda_j}=&\lambda_jC_j(\br);\nn
\sum_j\lambda_j C_j(\br)\, C_j(\brp)=&\pi R^2\delta^{(2)}(\br-\brp).
\end{align}
Using the identity, 
\be
\label{pmu_rhoH}
\pnu\varrho_H=\pmu\ket{K_0}\bra{K_0}+\ket{K_0}\pmu\bra{K_0},
\ee
in Eq.~(\ref{H_SLDmuK0}), then taking the inner product of the latter with the bra $\pnu\bra{K_0}$ and applying the first of identities (\ref{Disk_identities}) repeatedly, and finally substituting the result into Eq.~(\ref{H_QFIpert2}) yields the following expression for the QFI matrix elements:
\ba
\label{QFI4}
H_{\mu\nu}&=4\epsilon^2\Re\sum_{i,j} {\lambda_jC_j(\br_0)\pnu\braket{K_0}{\lambda_i}\over\lambda_i+\lambda_j}\nn
                  &\times [\bra{\lambda_i}\pmu\ket{K_0}\lambda_jC_j(\br_0)+\lambda_iC_i(\br_0)\pmu\braket{K_0}{\lambda_j}].
\end{align}
Note that because of the factor $\lambda_j$ in the numerator, the sum over $j$ is automatically restricted to being over eigenstates in the support of the disk SPDO. This allows the sum over $i$ to be over {\em all} states, including those that span the null space of the SPDO, indexed by $i\in\cN$, and those that are over its support, $i\in\cS$. The double sum in Eq.~(\ref{QFI4}) when restricted to the null subspace for the $i$ sum for which $\lambda_i=0$, reduces to the value
\ba
\label{Nij}
&\pnu\bra{K_0}\sum_{i\in \cN}\ketbra{\lambda_i}{\lambda_i}\pmu\ket{K_0}\sum_j\lambda_jC_j^2(\br_0)\nn
&=\pi R^2\delta^{(2)}(0)\pnu\bra{K_0}\cI_N\pmu\ket{K_0},
\end{align}
where we used the first of the identities (\ref{Disk_identities}) and the null-subspace identity operator, $\cI_N$, given by Eq.~(\ref{cIN}), to arrive at the latter expression. Adding to this contribution the contribution from the support subspace of the SPDO for both the $i$ and $j$ sums in Eq.~(\ref{QFI4}) then yields
\ba
\label{QFI5}
H_{\mu\nu}&=2\epsilon\,\Re \pnu\bra{K_0}\cI_N\pmu\ket{K_0}\nn 
&+4\epsilon^2\Re\sum_{i,j\in \cS} {\lambda_jC_j(\br_0)\pnu\braket{K_0}{\lambda_i}\over\lambda_i+\lambda_j}\nn
&\times [\bra{\lambda_i}\pmu\ket{K_0}\lambda_jC_j(\br_0)+\lambda_iC_i(\br_0)\pmu\braket{K_0}{\lambda_j},
\end{align}
in which we regularized the divergent $\delta^{(2)}(0)$ as simply $1/A_H$, with $A_H=\pi \delta_0^2$ being the area of the hole.

\end{document}